\documentclass[11pt]{article}
\usepackage[utf8]{inputenc}
\usepackage{amsthm, amsmath, amssymb}
\usepackage{lineno, cite, enumerate}
\usepackage{mathtools, framed}
\usepackage{xurl}

\usepackage{tikz}
\usetikzlibrary{shapes.geometric, arrows}
\tikzstyle{state} = [thick, circle, minimum width=1.cm, text centered, draw=black]
\tikzstyle{cat} = [thick, rectangle, rounded corners, minimum width=8cm, text centered, draw=black]
\tikzstyle{cat1} = [thick, rectangle, rounded corners, minimum width=4cm, text centered, draw=black]
\tikzstyle{arrow} = [thick,->,>=stealth]
\tikzstyle{arrow1} = [thick,dashed,->,>=stealth]

\newtheorem{theorem}{Theorem}
\newtheorem{remark}[theorem]{Remark}
\newtheorem{example}[theorem]{Example}
\numberwithin{equation}{section}
\numberwithin{theorem}{section}

\def\R{\mathbb R}
\def\N{\mathbb N}

\title{Population growth in discrete time: a renewal equation oriented survey}

\author{B. Boldin$^{1}$, O. Diekmann$^{2,*}$, J.A.J. Metz $^{3, 4, 5, 6, 7}$ }

\begin{document}

\maketitle

\noindent{} 1. Faculty of Mathematics, Natural Sciences and Information Technologies, University of Primorska, Glagolja\v{s}ka 8, SI-6000 Koper, Slovenia

\noindent{} 2. Department of Mathematics, University of Utrecht, P.O. Box 80010, 3580 TA Utrecht, The Netherlands

\noindent{} 3. Exploratory Modeling of Human-natural Systems Research Group, Advancing Systems Analysis Program, 
International Institute of Applied Systems Analysis (IIASA), A-2361 Laxenburg, Austria

\noindent{} 4.  Institute of Biology, Leiden University, P.O. Box 9512, 2300RA Leiden, The Netherlands 

\noindent{} 5. Mathematical Institute, Leiden University, P.O. Box 9512, 2300RA Leiden, The Netherlands 

\noindent{} 6. Netherlands Centre for Biodiversity, Naturalis, P.O. Box 9517, 2300RA Leiden, The Netherlands

\noindent{} 7. Complexity Science and Evolution Unit, Okinawa Institute of Science and Technology Graduate University (OIST), Onna 904-0495, Japan

\noindent{} * Corresponding author. E-mail: O.Diekmann@uu.nl
\\

\thanks{Dedicated to Jim Cushing, on the occasion of his 80th birthday, and to Horst Thieme, on the occasion of his 
75th birthday, in appreciation of their wonderful contributions to building and analysing structured-population models.}

\begin{framed}
\footnotesize{
\noindent This is the Author Version of an article accepted for publication in Journal of Difference Equations and Applications. 
The final version will be available at https://www.tandfonline.com/toc/gdea20/current.}
\end{framed}

\begin{abstract}
Traditionally, population models distinguish individuals on the basis of their current state. Given a distribution, a discrete time model then specifies 
(precisely in deterministic models, probabilistically in stochastic models) the population distribution at the next time point. The renewal equation alternative concentrates on newborn individuals and the model specifies the production of offspring as a function of age. This has two advantages: (i) as a rule, there are far fewer birth states 
than individual states in general, so the dimension is often low; (ii) it relates seamlessly to the next-generation matrix and the basic reproduction number. 
Here we start from the renewal equation for the births and use results of Feller and Thieme to characterise the asymptotic large time behaviour. Next we 
explicitly elaborate the relationship between the two bookkeeping schemes. This allows us to transfer the characterisation of the large time behaviour to traditional 
structured-population models.
\end{abstract}

\textit{Keywords}: structured-population model, renewal equation, growth rate, basic reproduction number, next-generation matrix, reproductive value, stable distribution

\section{Introduction}

In the pioneering paper \cite{cushing1994}, Jim Cushing and Zhou Yicang
\begin{enumerate}[(i)]
\item defined the net reproductive number (aka the basic reproduction number and in the present paper denoted by ${\mathcal R}_0$) in the context of linear discrete time 
population models, while highlighting its interpretation as the expected lifetime number of offspring, 
\item showed that the stability of the extinction state (i.e., the trivial steady state) is governed by the sign of ${\mathcal R}_0 - 1$ (where, in the case of nonlinear models, 
${\mathcal R}_0$ refers to the linearised model), 
\item demonstrated, by way of examples, that often ${\mathcal R}_0$ is much easier to determine than the real time growth rate (in the present paper denoted by $\rho$).
\end{enumerate}

By making systematic use of Perron-Frobenius theory, Chi-Kwong Li and Hans Schneider streamlined the proofs and extended these results in 
\cite{li2002}. For far-reaching generalisations we refer to recent papers of Horst Thieme, in particular \cite{thieme2020, thieme2023}. 
And for similar continuous time results in the context of epidemic models to \cite{thieme2018, diekmann1990, diekmann2010, diekmann2013}. Also see 
\cite{cushing2016}.

Over the years (and in close collaboration with various other authors) we have developed theory for continuous time structured-population 
models \cite{metz1986, diekmann1998, diekmann2001,diekmann2003}. These papers deal with rather general models and may, as a consequence, 
not be easily accessible. The aim of the present paper is to formulate the discrete time version of the linear theory in a reader-friendly fashion. 
To this end, we limit our attention to population models in which individuals are distinguished by a number of traits (e.g. species, age, size, spatial location, infection status), 
but such that the set of conceivable individual states (shortly, $i$-states) is finite.  

In our approach, the renewal equation (RE) for the birth rate takes centre stage. Once one solves the RE constructively, all other relevant quantities are given by 
explicit expressions in terms of the initial condition and the birth rate. As a consequence, we can deduce the asymptotic large time behaviour of every quantity of 
interest from the asymptotic large time behaviour of the birth rate. And to determine the latter, we can rely on Feller’s celebrated Renewal Theorem for scalar RE and on 
Thieme's generalisation for systems of RE (see Section XIII.10 in \cite{feller1968} and \cite{thieme1984}; continuous time results 
can be found in \cite{feller1971, crump1970} and \cite{desaporta2003}).

The key feature that makes this approach both attractive and efficient is that, as a rule, there are far fewer individual birth states than individual states in general. 
In particular, the RE is a scalar equation when all individuals are identical at birth, either in the literal sense (like, for instance, in the case of age as 
the only structuring trait) or in the stochastic sense, when states-at-birth have a fixed probability distribution (for example, in an epidemiological model 
where individuals begin their infected life (i.e. are born epidemiologically) as symptomatic with probability $p$ or 
asymptomatic with probability $1-p$, regardless of the $i$-state of the individual that infected them).

\begin{figure}[t]
\begin{center}
\begin{tikzpicture}[node distance=3cm and 3cm]
\node (C1) [cat] {\begin{tabular}{l}{\bf Structured-population model (SPM)} \\[2mm]$X(t) = (F+T)X(t-1), \ X \in \R^{n}$ \end{tabular}};
\node (C2) [cat, below of = C1] {\begin{tabular}{l}{\bf Renewal equation (RE)} \\[2mm]$B(t) = \sum_{s = 1}^{t}{K(s)B(t-s) + G(t)}, \ B \in \R^{m}$  \end{tabular}};
\node (C3) [cat1] [right = 5.2cm] at (C2){\begin{tabular}{l}{\bf RE asymptotic} \\  {\bf behaviour }\\[2mm] $\diamond$ Stable distribution $\Phi$ \\ $\diamond$ 
Reproductive values $\Psi$\end{tabular}};
\node (C4) [cat1] [above = 2.0cm] at (C3){\begin{tabular}{l}{\bf SPM asymptotic} \\  {\bf behaviour } \\[2mm] $\diamond$ Stable distribution $\Phi_{\rm r}$ \\ $\diamond$ 
Reproductive values $\Psi_{\rm r}$\end{tabular}};
\draw [arrow] (C1) -- node[anchor=west] {1.} (C2);
\draw [arrow] (C2) -- node[anchor=south] {2.} (C3);
\draw [arrow] (C3) -- node[anchor=west] {3.} (C4);
\end{tikzpicture}
\end{center}
\label{fig:diag}
\caption{Deducing the asymptotic large time behaviour in (linear discrete time) structured-population models via the corresponding renewal equation for the birth rate.}
\end{figure}
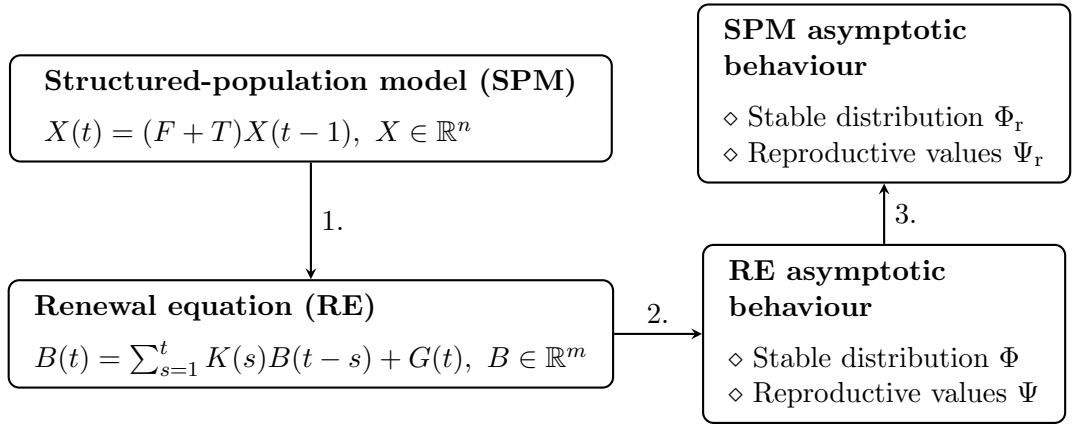

We begin by preparing the ground: in Section 2, we formulate the RE for the birth rate and determine the asymptotic large time behaviour using the theorems of 
William Feller (for scalar RE) and Horst Thieme (for systems of RE). In Section 3 we first show how classical discrete time structured-population models (SPM)
give rise to (typically much smaller) systems of RE for the birth rate and discuss how the two bookkeeping schemes relate to each other in terms of dynamics and 
next-generation matrices. Next we show how, after having deduced the asymptotic behaviour of solutions of the RE using the results of Section 2, one can 
determine the asymptotic $i$-state distribution and the reproductive values in SPM from the ones obtained in the corresponding RE (see  Figure \ref{fig:diag}). 
   
In Section 6 of \cite{diekmann1998} and Section 3.2 of \cite{diekmann2003} it was shown that, for the special situation that ${\mathcal R}_0 = 1$, not only the spectral 
equivalence $\rho = 1$ holds, but that one can also compute the asymptotic distribution (i.e. the right eigenvector) in the real time setting from that in the generation 
bookkeeping framework. In Section 4 we supplement this result with a corresponding result for the reproductive values (or, in other words, for the left eigenvector). 

In \cite{diekmann2003} it was demonstrated that for a large class of nonlinear models the condition ${\mathcal R}_0 = 1$ arises when looking for steady states. 
This makes the results about connecting the eigenvectors useful for nonlinear models as well.

In Appendix we present a brief outline of continuous time analogues of various equations and identities derived in Sections 2, 3 and 4. 

\section{The renewal equation}

The first step in the formulation of renewal equations is the identification of states-at-birth. We may interpret this literally in the sense of determining all 
the $i$-states in which individuals may begin their lives. For example, think of a model where individuals are structured by age $a$ and size $s$ and where 
the set of $i$-states is $\{(a,s):  a = \{1, \hdots, k\}, s = \{{\rm small}, {\rm large}\}\}.$ Assuming that newborns may be either small or large, there are two states-at-birth: 
$(1, {\rm small})$ and $(1, {\rm large}).$ However, when small and large offspring is produced at a fixed ratio $p:(1-p)$ (independently of the $i$-state of the parent) 
we can exploit the fact that there is a single state-at-birth in the stochastic sense: all individuals are born with the same probability distribution of their $i$-state. Another example is 
an epidemiological model where individuals may begin their infected life (i.e. are born epidemiologically) as either asymptomatic or symptomatic. There are again 
two states-at-birth in the literal sense, but when new cases arise at a fixed ratio there is but one state-at-birth in the stochastic sense. 

Both interpretations of the term state-at-birth are valid. However, while the literal interpretation may be a natural starting point, understanding the 
term in the stochastic sense allows us to formulate RE systems of minimal size (we elaborate this point in Section 3). 

We begin by considering the case where all individuals are identical at birth, i.e. are born with the same probability distribution of their $i$-state.

\subsection{The scalar equation}

Imagine that immediately after the yearly breeding season a census is made of the subpopulation of newborn females. Let $b(t)$ denote that number 
in year $t$. If we ignore density dependence and adopt a deterministic point of view at the population level (i.e. we assume that the numbers are so large that it 
makes sense to focus on expected values and to ignore the effects of demographic stochasticity) then 

\begin{linenomath*}
\begin{equation}
b(t) = \sum_{s = 1}^{\infty}{k(s) b(t-s)},
\label{REscalar}
\end{equation}
\end{linenomath*}
where $k(s)$ is the expected number of daughters born to a female in year $s$ after her own birth.  We 
consider the sequence $\{k(t)\}_{t = 1}^{\infty}$ as the basic model ingredient and make the following assumptions: 

\begin{enumerate}
\item [A1.] $k(t) \geq 0$ for all $t \in {\mathbb N}$ with strict inequality for at least one value of $t$.
\item [A2.] $\sum_{t = 1}^{\infty}{k(t)} < \infty$.
\end{enumerate}

Note that 
\begin{linenomath*}
\begin{equation}
{\mathcal R}_0 = \sum_{t = 1}^{\infty}{k(t)}
\label{R0scalar}
\end{equation}
\end{linenomath*} is the {\em basic reproduction number}, i.e. the expected lifetime number of (female) offspring. 

Equation \eqref{REscalar} is linear and translation invariant, so we expect it to have geometric solutions. If we substitute the Ansatz 
\begin{linenomath*}
\begin{equation}
b(t) = z^{t}
\label{ansatz}
\end{equation}
\end{linenomath*}
into \eqref{REscalar} and divide both sides by $z^t$ we obtain the Euler-Lotka characteristic equation 
\begin{linenomath*}
\begin{equation}
1 = \sum_{s = 1}^{\infty}{k(s)z^{-s}} =: \bar{k}(z) 
\label{ELscalar}
\end{equation}
\end{linenomath*}
for the unknown $z$. Note that $\bar{k}(z)$ is the (one-sided) z-transform of  $\{k(t)\}_{t = 0}^{\infty}$ (with $k(0) :=0$) and that the $z$-transform is the discrete time analogue 
of the Laplace transform. 

For real $z$, $\bar{k}$ is a strictly decreasing function with limit zero at infinity. So if $\bar{k}$ takes, for a real $z$, a value larger than one, 
equation \eqref{ELscalar} has precisely one real solution. We denote this solution by $\rho$ and observe that there are precisely three possibilities: 
\begin{linenomath*}
\begin{align*}
\begin{split}
 &  1 < \rho < {\mathcal R}_0, \\
 & 1 = \rho = {\mathcal R}_0, \\
 & {\mathcal R}_0 < \rho < 1
\end{split}
\end{align*}
\end{linenomath*}
whenever $k(s)>0$ for a value $s\geq 2$ (in the degenerate case that only $k(1)>0$ we have $\rho = {\mathcal R}_0 = k(1)$). To see this, note that 
\begin{enumerate}[(i)]
\item  $\bar{k}(1) = {\mathcal R}_0$, 
\item if ${\mathcal R}_0 >1$ then $\bar{k}({\mathcal R}_0) <1$, 
\item if ${\mathcal R}_0 <1$ then $\bar{k}({\mathcal R}_0)>1$ (possibly infinite). 
\end{enumerate}
If $z$ is complex then 
\begin{linenomath*}
\begin{equation*}
 |\bar{k}(z)| \leq \bar{k}(|z|).
 \end{equation*}
\end{linenomath*}
Consequently the solutions of \eqref{ELscalar} are contained within the closed disc of radius $\rho$. 

We say that $\{k(t)\}_{t = 1}^{\infty}$ is {\em periodic} with period $p>1$ if $k(s) = 0$ except when
$s$ is an integer multiple of $p$. One easily verifies that in that case $z$ is a solution of \eqref{ELscalar} if $z$ is the product of $\rho$ and a $p$-th root of unity. So for 
periodic $\{k(t)\}_{t = 1}^{\infty}$ there are non-real solutions of \eqref{ELscalar} on the circle of radius $\rho$ in ${\mathbb C}$.
We make a further assumption:

\begin{enumerate}
\item [A3.] $\{k(t)\}_{t = 1}^{\infty}$ is not periodic. 
\end{enumerate}

So far we focused on special solutions of \eqref{REscalar} of the form \eqref{ansatz}. Now, in the spirit of \cite{diekmann2008}, we prescribe the history of $b$ up to 
$t  =0$ by putting 
\begin{linenomath*}
\begin{equation}
b(\tau) = \theta(\tau), \ \ \ \tau = \hdots, -2, -1
\label{bhist}
\end{equation}
\end{linenomath*}
for a given function $\theta(\tau) \geq 0$ such that 
\begin{linenomath*}
\begin{align}
\begin{split}
g(t) &:= \sum_{s = t+1}^{\infty}{k(s)\theta(t-s)} =  \sum_{l = 1}^{\infty}{k(l+t)\theta(-l)}
\label{gscalar}
\end{split}
\end{align}
\end{linenomath*}
is finite for $t \in \N$. This allows us to rewrite \eqref{REscalar} in the form 
\begin{linenomath*}
\begin{align}
b(t) &= \sum_{s = 1}^{t}{k(s)b(t-s)} + g(t), \ \ \ t \in \N, 
\label{REbgk}
\intertext{
with the convolution product interpreted as zero for $t = 0$, i.e.,}
b(0) &= g(0).
\label{REbgini}
\end{align}
\end{linenomath*}
Let $\bar{b}$ and $\bar{g}$ denote the (one-sided) $z-$transforms of, respectively, $\{b(t)\}_{t = 0}^{\infty}$ and $\{g(t)\}_{t = 0}^{\infty}$, i.e. 
\begin{linenomath*}
\begin{equation*}
%\label{barG}
\bar{b}(z) = \sum_{s = 0}^{\infty}{b(s) z^{-s}}, \ \ \ \bar{g}(z) = \sum_{s = 0}^{\infty}{g(s) z^{-s}.}
\end{equation*}
\end{linenomath*}
The $z$-transform 
has the useful property that it turns a convolution product into an ordinary product. 
So in terms of the $z$-transforms, the convolution equation \eqref{REbgk} can be written as 
\begin{linenomath*}
\begin{equation*}
\bar{b} = \bar{k}\bar{b} + \bar{g}.
\end{equation*}
\end{linenomath*}
If follows that 
\begin{linenomath*}
\begin{equation*}
\bar{b} = (1-\bar{k})^{-1}\bar{g}
\end{equation*}
\end{linenomath*}
and next, by taking the inverse $z$-transform, that 
\begin{linenomath*}
\begin{equation*}
b(t) = \frac{1}{2\pi i} \oint (1-\bar{k}(z))^{-1}\bar{g}(z)z^{t-1}dz
\end{equation*}
\end{linenomath*}
where the closed contour encircles the origin counter-clockwise at a distance great enough to enclose all singularities of the integrand. The observations made above 
concerning the solutions of \eqref{ELscalar} now suggest that $b(t) \approx \rho^{t}$ for $t \to \infty$. Theorem \ref{FRT} in Section XIII.10 of \cite{feller1968} gives a far more 
informative characterisation of the asymptotic large time behaviour of $b$.

\begin{theorem} (Feller's Renewal Theorem)
\label{FRT}
Consider the renewal equation \eqref{REbgk} and assume that 
 A1, A2 and A3 hold. Furthermore, assume that $g(t)\geq 0$  for all 
$t \in \N$ and that $\sum_{t= 0}^{\infty}{g(t)} < \infty$. For ${\mathcal R}_0 >1$ let $\rho >1$ denote the unique real solution of the Euler-Lotka equation \eqref{ELscalar}.
\begin{enumerate}[(i)]
\item If ${\mathcal R}_0 <1$ then $b(t) \to 0$ for $t \to \infty$ and 
\begin{linenomath*}
$$\sum_{t = 0}^{\infty}{b(t)} = \frac{\bar{g}(1)}{1-{\mathcal R}_0}.$$
\end{linenomath*}
\item If ${\mathcal R}_0 = 1$ then for $t \to \infty$
\begin{linenomath*}
$$ b(t) \to \frac{\bar{g}(1)}{\sum_{s = 1}^{\infty}{s k(s)}}.$$
\end{linenomath*}
\item  If ${\mathcal R}_0 >1$ then for $t \to \infty$
\begin{linenomath*}
$$ \rho^{-t}b(t) \to \frac{\bar{g}(\rho)}{\sum_{s  =1}^{\infty}{sk(s)\rho^{-s}}}.$$
\end{linenomath*}
\end{enumerate}
\end{theorem}

\begin{remark} 
To facilitate comparison with Feller's formulation in \cite{feller1968}, we present the notation-translation table: 
\textup{
\begin{center}
\begin{tabular}{ |c|c| } 
\hline
Here & Feller \cite{feller1968} \\ 
\hline \hline
b & v \\ 
k & f \\
g & b \\
 \hline
\end{tabular}.
\end{center}}
More importantly, note that Feller uses the generating function rather than the z-transform. If $\tilde{k}$ denotes the generating function of 
$\{k(t)\}_{t = 0}^{\infty}$ then
\begin{linenomath*}
\begin{equation*}
\tilde{k}(s) = \bar{k}(s^{-1}). 
%\ \ \ \mbox{and} \ \ \ \tilde{k}'(s) = -s^{-2}\bar{k}'(s^{-1}).
\end{equation*}
\end{linenomath*}
\end{remark}
If  \eqref{gscalar} holds we deduce from the second identity that 
\begin{linenomath*}
\begin{subequations}
\begin{align}
\bar{g}(\rho) &= \sum_{l = 1}^{\infty}{c(l) \theta(-l)} 
\label{barGc}
\intertext{with} 
c(l) &:= \sum_{s = 0}^{\infty}{\rho^{-s}k(l+s)}.
\label{c}
\end{align}
\end{subequations}
\end{linenomath*}
The coefficients $c(l)$ thus tell us how the various components of the initial condition \eqref{bhist} contribute to the ultimate geometric population growth with 
multiplication factor $\rho$ 
when ${\mathcal R}_0 >1$ (for the time being, we do not normalise these contributions). 

We conclude that Theorem \ref{FRT} gives a complete description of the large time behaviour of solutions of the renewal equation \eqref{REbgk} when we prescribe 
as an initial condition the history of $b$ up to time zero as in \eqref{bhist}.

\subsection{Systems of renewal equations}

Suppose that individuals begin their lives in one of $m>1$ states-at-birth. 
The renewal equation now reads 
\begin{linenomath*}
\begin{equation}
B(t) = \sum_{s = 1}^{\infty}{K(s) B(t-s)},
\label{REsystems}
\end{equation}
\end{linenomath*}
where the components $b_j(t)$ of the vector $B(t) \in {\mathbb R}^m$ denote the number of females born at time $t$ with state-at-birth $j$ and 
$K(s)$ is a positive $m\times m$ matrix with elements 
\begin{linenomath*}
\begin{align*}
k_{ij}(s) =& \  \mbox{\it the expected number of daughters with state-at-birth $i$ produced by a} \\
& \ \mbox{\it female that was herself born $s$ time units ago with state-at-birth $j$.}
\end{align*}
\end{linenomath*}

\begin{remark} 
A matrix $A$ is called positive if $a_{ij} \geq 0$ for all $i$ and $j$. In this case, we write $A \geq 0$. We write   
$A > 0$ when $A \geq 0$ and $A \ne 0$.  
A matrix $A$ is called strictly 
positive if $a_{ij} > 0$ for all $i$ and $j$. In this case, we write $A \gg 0$. In Appendix we use that a matrix $A$ is called positive-off-diagonal if $a_{ij} \geq 0$ for 
 $i \ne j$ and $a_{ii} \in \R$ \cite{batkai2017}.
\end{remark}

The sequence of matrices $K = \{K(t)\}_{t = 0}^{\infty}$ (with $K(0):= 0$) forms a kernel. We assume 
\begin{enumerate}
\item [A4.] $K(t) \geq 0$ for all $t \in {\mathbb N}$ with $K(t)>0$  for at least one $t\in \N$.
\item [A5.] $ \sum_{t = 0}^{\infty}{\|K(t)\|} < \infty$.
\end{enumerate}

\begin{remark}
In a population dynamical context, it is preferable to equip ${\mathbb R}^m$ with the $l_1$-norm, i.e., to define 
$\|X\| = \sum_{j = 1}^{m}{|x_j|}.$
In A5, we then use the corresponding operator norm, 
\begin{linenomath*}
$$\|K(t)\| = \max_{1 \leq j \leq m}{\sum_{i = 1}^{m}{|K(t)_{ij}|}}.$$
\end{linenomath*}
\end{remark}

If we now look for solutions of the RE \eqref{REsystems} in the form  
$ B(t) = z^{t}\Phi$
for some $m$-vector $\Phi$ we observe that the Euler-Lotka equation takes the form 
\begin{linenomath*}
\begin{align}
r_{\sigma}(\bar{K}(z)) = 1, 
\label{ELsystems}
\end{align}
\end{linenomath*}
where $r_\sigma$ denotes the {\it spectral radius} and where, as before, 
\begin{linenomath*}
\begin{equation}
\bar{K}(z) = \sum_{s = 0}^{\infty}{K(s)z^{-s}}.
\label{Kbar}
\end{equation}
\end{linenomath*}
We call $\bar{K}(1)$ the next-generation matrix and define 
 \begin{linenomath*}
 \begin{equation}
 {\mathcal R}_0 := r_{\sigma}(\bar{K}(1)).
 \label{R0systems}
  \end{equation}
\end{linenomath*}
Note (by exploiting the fact that the spectral radius of $\bar{K}(z)$ is a decreasing function of $z$) that the Euler-Lotka equation \eqref{ELsystems} has a unique real solution 
$\rho$ when ${\mathcal R}_0$ is larger than or equal to one. In that case, we necessarily have 
$ 1 < \rho < {\mathcal R}_0$ or  $1 = \rho = {\mathcal R}_0.$  Furthermore, $\bar{K}(1)$ and $\bar{K}(\rho)$ are positive matrices, and therefore for both matrices 
the spectral radius is a dominant eigenvalue with a positive corresponding eigenvector \cite{berman1994, batkai2017}.

The key reference concerning the asymptotic behaviour of the solutions of the renewal equation \eqref{REsystems} is now \cite{thieme1984}. 
In order to present Thieme's renewal theorem, we again prescribe the history of $B$ up to $t = 0$, 
\begin{linenomath*}
\begin{equation}
B(\tau) = \Theta(\tau), \ \ \ \tau = \hdots, -2, -1
\label{Bhistsys}
\end{equation}
\end{linenomath*}
for given positive $\Theta(\tau) \in \R^m$ such that 
\begin{linenomath*}
\begin{align}
G(t) &:= \sum_{s = t+1}^{\infty}{K(s)\Theta(t-s)} =  \sum_{l = 1}^{\infty}{K(l+t)\Theta(-l)} 
\end{align}
\end{linenomath*}
defines a vector with finite components for all $t \in \N$. This allows us to write 
 \eqref{REsystems} as
\begin{align}
B(t) &= \sum_{s = 1}^{t}{K(s)B(t-s)} + G(t),  \ \ \ t \in \N
\label{REBKGsys}
\end{align}
with the convention
\begin{linenomath*}
\begin{align}
 B(0) &= G(0).
\end{align}
\end{linenomath*}
Next, we introduce the resolvent of the kernel $K$. The resolvent $R = \{R(t)\}_{t = 0}^{\infty}$ is defined by the equation 
\begin{linenomath*}
\begin{equation*}
\label{resolvent}
R = K \ast R + K = R \ast K + K
\end{equation*}
\end{linenomath*}
where $\ast$ denotes the convolution product, i.e.
\begin{linenomath*}
\begin{equation}
K \ast R  (t) := \sum_{s = 0}^{t}{K(s)R(t-s)}.
\label{conv}
\end{equation}
\end{linenomath*}
In our case $K(0) = 0$ and consequently $R(0) = 0$, meaning that we can write \eqref{conv} as 
\begin{linenomath*}
\begin{equation*}
\label{conv1}
K \ast R (t) = \sum_{s = 1}^{t-1}{K(s)R(t-s)}.
\end{equation*}
\end{linenomath*}

The results in \cite{thieme1984} concern a rather general setting in terms of positive linear operators on ordered Banach spaces. When dealing with spaces of 
functions defined on non-compact domains, it often helps to define positivity in terms of a special element $w$ of the positive cone, with $w$ possibly tending to 
zero for the argument tending to infinity. Here, however, we only deal with the standard positive cone in ${\mathbb R}^m$ and when invoking the results of 
\cite{thieme1984} we simply take the m-vector $w= [1, \hdots, 1]^{T}$.

We make further assumptions
\begin{enumerate}
\item [A6.] There exists $t_0 \in {\mathbb N}$ such that $R(t_0)\gg0$.
\item [A7.] There exist $t_1, t_2 \in {\mathbb N}$ such that their greatest common divisor is one and $R(t_i)>0$ ($i = 1,2$).
\end{enumerate}

\begin{theorem}(Thieme's Renewal Theorem). 
\label{TRT} Consider the renewal equation  \eqref{REBKGsys} and 
assume that A4, A5, A6 and A7 hold. Furthermore, assume that there exists $M >0$ such that 
$\|G(t)\| \leq M$ for all $t \in \N$.  Assume that ${\mathcal R}_0 \geq 1$. The Euler-Lotka equation \eqref{ELsystems} has a unique real solution $\rho$ with either 
$1< \rho < {\mathcal R}_0$ or $\rho = 1 = {\mathcal R}_0$. Furthermore let 
$\Phi$, with $\|\Phi\| = 1$ and $\Psi$, with $\Psi \Phi = 1$, denote, 
respectively, the right and the left normalised positive eigenvector of $\bar{K}(\rho)$ corresponding to eigenvalue 1.  

Then for $t \to \infty$
\begin{linenomath*}
\begin{subequations}
\begin{align}
&\rho^{-t}B(t) \to c\Phi 
\intertext{
with}
&c = \Psi \frac{\bar{G}(\rho)}{c_0} \ \ \ \mbox{and} \ \ \ c_0 = \Psi \Big(\sum_{s= 1}^{\infty}{s\rho^{-s}K(s)\Phi}\Big).
\label{ThiemeC}
\end{align}
\end{subequations}
\end{linenomath*}
\end{theorem}

\begin{remark} 
\begin{enumerate}[(i)]
\item The following table aims to facilitate comparison with Thieme's results in \cite{thieme1984}: 
\textup{
\begin{center}
\begin{tabular}{ |c|c| } 
\hline
Here & Thieme \cite{thieme1984} \\ 
\hline \hline
$B$ & $u$ \\ 
$K$ & $A$ \\
$G$ & $\bar{u}$ \\
 \hline
\end{tabular}.
\end{center}}
\item Note that one can normalise $\Psi$ in any way one wants, since in \eqref{ThiemeC} $\Psi$ is a factor in both the numerator and the denominator. Our choice,  
 $\Psi \Phi = 1$, is handy in the context of many applications.
\end{enumerate}
\end{remark}

\section{Structured-population models}

When density dependence is ignored and the set of $i$-states is finite, a discrete time structured-population model leads to the system 
of linear recursion relations
\begin{linenomath*}
\begin{equation}
\label{recurrsion}
X(t) = (F+T)X(t-1). 
\end{equation}
\end{linenomath*}
Here the components $x_j(t)$ of the vector $X(t) \in \R^n$ correspond to the 
density of females with $i$-state $j$ at time $t$, while $F$ and $T$ are 
positive $n \times n$ matrices that describe, respectively, reproduction and survival $\&$ development, cf. \cite{cushing1998, caswell2000}. 
More precisely, $F = (f_{ij})_{i,j = 1}^{n}$ and $T = (t_{ij})_{i,j = 1}^{n}$ where 
\begin{linenomath*}
\begin{align*}
f_{ij} = \  &\mbox{\it the expected number of daughters with $i$-state $i$,} \\
& \mbox{\it produced by a female with $i$-state j} \\
\intertext{and}
t_{ij} = \ &\mbox{\it the probability that an individual with $i$-state $j$ } \\
& \mbox{\it is alive and has $i$-state $i$ one time unit later. }
\end{align*}
\end{linenomath*}
Since no individual is immortal we assume that $r_{\sigma}(T)<1$. 

The aim of this section is twofold: (i) we show how the recurrence \eqref{recurrsion} and the renewal equation \eqref{REsystems} relate to each other and 
(ii) we discuss how the next-generation matrix and the basic reproduction number corresponding to \eqref{recurrsion} are connected with \eqref{R0systems}. 
We begin with the former. 

Suppose there are $m$ states-at-birth in the literal sense ($1 \leq m \leq n$). There exist a positive $n \times m$ matrix $V$ (normalised such that all $m$ columns of 
$V$ have $l_1$-norm equal to one) and a positive $m \times n$ matrix $U$ such that 
\begin{linenomath*}
\begin{equation}
F = VU.
\label{FVU}
\end{equation}
\end{linenomath*}

In particular, if the states-at-birth are $j_1, \hdots, j_m \in \{1, \hdots, n\}$ we define $V$ as the matrix with columns 
$e_{j_1}, \hdots, e_{j_m}$ and $U$ as the fertility matrix obtained by taking the rows $j_1, \hdots, j_m$ from $F$. 

Directly from the interpretation we discover the first relation, namely,  
\begin{linenomath*}
\begin{equation}
B(t) = UX(t-1).
\label{BUX}
\end{equation}
\end{linenomath*}
Using the generation expansion we then deduce from \eqref{recurrsion}, \eqref{FVU} and \eqref{BUX} that 
\begin{linenomath*}
\begin{align*}
X(t) &= VB(t) + TX(t-1) \\
&= VB(t) + TVB(t-1) + T^2X(t-2) \\
&= \hdots, 
\end{align*}
\end{linenomath*}
leading in the limit to the second relation 
\begin{linenomath*}
\begin{equation}
X(t) = \sum_{s = 0}^{\infty}{T^sVB(t-s)}.
\label{XB}
\end{equation}
\end{linenomath*}
The two equations \eqref{BUX} and \eqref{XB} can both be understood on the basis of their interpretation and together they provide a complete description of the 
dynamics. By substituting \eqref{XB} into \eqref{BUX} one obtains \eqref{REsystems} with 
\begin{linenomath*}
\begin{equation}
K(s) = UT^{s-1}V.
\end{equation}
\end{linenomath*}
Likewise one recovers \eqref{recurrsion} from \eqref{BUX} and \eqref{XB} by splitting off the $s = 0$ term in \eqref{XB} to 
obtain 
\begin{linenomath*}
\begin{equation}
X(t) = VB(t) + TX(t-1)
\end{equation}
\end{linenomath*}
and next use \eqref{BUX} and \eqref{FVU}.

The two equations \eqref{BUX} and \eqref{XB} relate the time course of $B$ to that of $X$ and vice versa, when defined for all $t$, not just for $t \geq 0$. 
To deal with the initial value problem for $X$, we can rewrite \eqref{recurrsion} in the variation-of-constants form 
\begin{linenomath*}
\begin{equation}
X(t) = T^{t}X(0) + \sum_{s= 1}^{t}{T^{s-1}FX(t-s)}.
\label{REx}
\end{equation}
\end{linenomath*}
By acting on both sides of this identity with $F$ we obtain 
\begin{linenomath*}
\begin{equation}
Y(t) = FT^{t}X(0) + \sum_{s = 1}^{t}{FT^{s-1}Y(t-s)}
\label{REy}
\end{equation}
\end{linenomath*}
where 
\begin{linenomath*}
\begin{equation*}
Y(t) = FX(t).
\end{equation*}
\end{linenomath*}
Once we solve the RE \eqref{REy}, we can rewrite \eqref{REx} as the explicit expression 
\begin{linenomath*}
\begin{equation}
\label{xy}
X(t) = T^{t}X(0) + \sum_{s = 1}^{t}{T^{s-1}Y(t-s)}.
\end{equation}
\end{linenomath*}
Here $Y$ takes values in $\R^n$, but whenever $F$ has the form \eqref{FVU}, we know that 
\begin{linenomath*}
\begin{equation}
Y(t) = VB(t)
\label{YVB}
\end{equation}
\end{linenomath*}
for a function $B$ taking values in $\R^m$. We can then rewrite \eqref{REy} in the form 
\begin{linenomath*}
\begin{equation}
B(t) = UT^tX(0) + \sum_{s = 1}^{t}{UT^{s-1}VB(t-s)}, 
\label{Bsys}
\end{equation}
\end{linenomath*}
that is, we get \eqref{REBKGsys} with 
\begin{linenomath*}
\begin{subequations}
\label{KGsys}
\begin{align}
K(s) &= UT^{s-1}V, 
\label{Ksys}
\\
G(s) &= UT^sX(0).
\label{Gsys}
\end{align}
\end{subequations}
\end{linenomath*}

Note that when the rank of $U$ is not maximal, there are fewer states-at-birth in the stochastic sense than there are in the literal sense. In such a case, a further reduction 
in RE system size is possible with a different choice of $U$ and $V$ (see one such example below). 

Hence, if the conditions of Theorem \ref{TRT} (or Theorem \ref{FRT} in the case where \eqref{Bsys} amounts to a scalar equation) are satisfied, we can first deduce the asymptotic behaviour 
of $B(t)$ from that theorem and then use \eqref{xy}, with $Y$ given by \eqref{YVB}, to determine the large time behaviour of the population state $X(t)$. We work out the details 
of this procedure in the two subsections that follow. But first, we discuss how the next-generation matrix and the basic reproduction number corresponding to \eqref{recurrsion} are connected with \eqref{R0systems}.

For $t \in \N$, the $(i,j)$-th element of $T^{t}$ is the probability that an individual with $i$-state $j$ is alive $t$ units of time later and has then $i$-state $i$. 
If $r_{\sigma}(T)<1$ then $(I-T)^{-1}$ exists and the $(i,j)$-th element of 
\begin{linenomath*}
$$(I-T)^{-1} = \sum_{t = 0}^{\infty}{T^t}$$
\end{linenomath*}
equals the expected time an individual starting with $i$-state $j$ will spend in $i$-state $i$. 

The matrix 
\begin{linenomath*}
\begin{equation}
{\mathcal L}_{\rm L} = F(I-T)^{-1}
\label{NGM}
\end{equation}
\end{linenomath*}
is the {\it next-generation matrix {\rm (NGM)} with large domain}, with the $(i,j)$-th element of ${\mathcal L}_{\rm L}$ specifying the expected number of future offspring with $i$-state $i$ produced by an  
 individual presently having $i$-state $j$. Often one defines 
\begin{linenomath*}
\begin{equation}
{\mathcal R}_0 = r_{\sigma}({\mathcal L}_L).
\label{R0L}
\end{equation}
\end{linenomath*}

The elements of the NGM with large domain \eqref{NGM} specify the expected numbers of future offspring of individuals for all $i$-states. 
But, with the factorisation of $F$ as described directly after \eqref{FVU}, we can focus on the true NGM 
\begin{linenomath*}
\begin{equation}
{\mathcal L} = U(I-T)^{-1}V. 
\label{NGMs}
\end{equation}
\end{linenomath*}
We invite the reader to verify that the $(i,j)$-th element of ${\mathcal L}$ gives the expected lifetime number of offspring with state-at-birth $i$ produced by a 
newborn individual with state-at-birth $j$. 

Note that when the rank of $U$ is smaller than $m$ (which happens whenever there are fewer states-at-birth in the stochastic sense than 
there are in the literal sense), a further reduction to the NGM {\it with small domain} ${\mathcal L}_{\rm S}$ is possible by altering $U$ and $V$ 
(see one example below and also \cite{diekmann2010} for the derivation of NGM with small and large domains in the context of epidemiological models 
in continuous time). 

One easily verifies that 
\begin{enumerate}[(i)]
\item 
the NGM with large domain \eqref{NGM} and the NGM \eqref{NGMs} have the same non-zero eigenvalues and therefore \eqref{R0L} amounts to 
\begin{linenomath*}
\begin{equation}
{\mathcal R}_0 = r_{\sigma}({\mathcal L}). 
\end{equation}
\end{linenomath*}
(And when further reduction to ${\mathcal L}_{\rm S}$ is possible then also ${\mathcal R}_0 = r_{\sigma}({\mathcal L}_{\rm S}).$)
\item 
If $\Phi$ is a right eigenvector of ${\mathcal L}$ corresponding to a non-zero eigenvalue $\lambda$ then $\tilde{\Phi} = V\Phi$ is a right eigenvector of 
${\mathcal L}_{\rm L}$ corresponding to eigenvalue $\lambda$. 
\item If $\tilde{\Psi}$ is a left eigenvector of ${\mathcal L}_{\rm L}$ corresponding to a non-zero eigenvalue $\lambda$ then $\Psi = \tilde{\Psi}V$ 
is a left eigenvector of ${\mathcal L}$ corresponding to eigenvalue $\lambda$. 
\end{enumerate}
Furthermore, when $r_{\sigma}(T)<1$ we deduce from \eqref{Kbar} and \eqref{Ksys} that 
\begin{linenomath*}
\begin{equation}
\bar{K}(1) = U(I-T)^{-1}V.
\end{equation}
\end{linenomath*}
That is, when $U$ and $V$ are defined as described after \eqref{FVU} then $\bar{K}(1)$ is the NGM and we recover \eqref{R0systems}.

\subsection{One state-at-birth}

The simplest situation arises when there is only one state-at-birth, either in the deterministic sense where all individuals are identical at birth (e.g. when age 
is the only structuring variable)
or in the stochastic sense where states-at-birth have a given probability distribution. In that case $F$ has one dimensional range. 

Then $F= VU$, where  $V \in {\mathbb R}^{n}$ (with $\|V\| = 1$)  
spans the range of $F$ and $U$ corresponds to  
the row $n$-vector (i.e. a $1\times n$ matrix) of fertility rates. In the notation of Section 2.1 we then define 
\begin{linenomath*}
$$b(t) = UX(t-1)$$
\end{linenomath*} 
and consider the scalar RE \eqref{REbgk} with 
\begin{linenomath*}
\begin{subequations}
\label{kg}
\begin{align}
\label{k} 
k(t) &= UT^{t-1} V 
\\ 
g(t) &= UT^{t}X(0). 
\label{g} 
\end{align}
\end{subequations}
\end{linenomath*}
Let's assume that $\{k(t)\}_{t = 1}^{\infty}$ and $\{g(t)\}_{t = 1}^{\infty}$ thus defined satisfy the conditions in Theorem \ref{FRT}. 

When $r_{\sigma}(T)<1$ we have 
\begin{linenomath*}
\begin{equation}
\label{barkU}
\bar{k}(z) = U(zI -T)^{-1} V
\end{equation}
\end{linenomath*}
for $|z|\geq 1$ and in particular 
\begin{linenomath*}
\begin{equation}
\label{R0U}
{\mathcal R}_0 = U(I-T)^{-1}V.
\end{equation}
\end{linenomath*}
If ${\mathcal R}_0 <1$ then $b(t) \to 0$ as $t \to \infty$ and consequently $X(t) \to 0$ as $t \to \infty$. 

Let's focus on the case ${\mathcal R}_0 \geq 1$. We can then compute the real time growth rate  $\rho$  as the unique solution of the Euler-Lotka equation 
\begin{linenomath*}
$$ \bar{k}(z) =1,$$
\end{linenomath*}
with necessarily either $\rho = 1 = {\mathcal R}_0$ or $1 < \rho < {\mathcal R}_0$. 
Since \eqref{barkU} can be written as 
\begin{linenomath*}
\begin{equation}
\bar{k}(z) = z^{-1}U(I -z^{-1}T)^{-1} V
\label{kbarscaled}
\end{equation}
\end{linenomath*}
we observe that the population growth rate $\rho$ is the value of $z$ for which both the generation and the real time process become stationary when we multiply all fertilities and 
all transition probabilities by $z^{-1}$. When $\rho >1$ the latter can be interpreted as introducing an additional, state-independent, death probability per time step 
$1-\rho^{-1}$. 

Then for $t \to \infty$ 

\begin{linenomath*}
\begin{equation*}
\label{Basym}
b(t) = c\rho^{t} + o(\rho^t)
\end{equation*}
\end{linenomath*}

for some constant $c$ and from \eqref{xy} and \eqref{YVB} it now follows that 
\begin{linenomath*}
\begin{align*}
\begin{split}
X(t) &= c \rho^t\sum_{s = 1}^{t}{\rho^{-s}T^{s-1}V} + o(\rho^t) \\ &= c\rho^t(\rho I - T)^{-1}V + o(\rho^t).
\end{split}
\end{align*}
\end{linenomath*}

That is, the solution $X(t)$ of \eqref{recurrsion} with $X(0) > 0$ grows, for large $t$, geometrically with rate $\rho$ while converging to 
the stable distribution 
\begin{linenomath*}
\begin{equation}
\Phi_{\rm r} = \frac{1}{\|(\rho I - T)^{-1}V\|}(\rho I- T)^{-1}V. 
\label{Phir_scalar}
\end{equation}
\end{linenomath*} 
Here the index {\rm r} refers to “real time” (as opposed to “generation”). 
By rewriting \eqref{Phir_scalar} as 
\begin{linenomath*}
\begin{equation}
\Phi_{\rm r} = \frac{1}{\|(I - \rho^{-1}T)^{-1}V\|}(I- \rho^{-1}T)^{-1}V
\label{Phir_scalar_2}
\end{equation}
\end{linenomath*} 
we can (with the $l_1$ norm) interpret the denominator 
as the life expectancy of a newborn individual when we introduce into the system an additional, state-independent, death probability per time step $1-\rho^{-1}$ (or, in other words, 
discount again and again the next year by a factor $\rho^{-1}$).

In the spirit of \eqref{c} we next ask: how do the components of $X(0)$ contribute to future population sizes, i.e. to the constant $c$? From \eqref{g} it follows that 
\begin{linenomath*}
\begin{align}
\label{barg}
\bar{g}(z) &= z U (zI-T)^{-1}X(0).
\end{align}
\end{linenomath*}
Now recall Theorem \ref{FRT}(iii) and conclude that the $j$-th component of the row vector 
\begin{linenomath*}
\begin{equation}
\Psi_{\rm r} = cU(\rho I-T)^{-1}
\end{equation}
\end{linenomath*}
specifies the (relative) contribution of individuals with $i$-state $j$ to future population size. Or, in the now generally accepted terminology introduced by Fisher \cite{fisher1958}, specifies the \textit{reproductive value} of state $j$. 

It is straightforward to check that 
\begin{enumerate}[(i)]
\item $F + T$ has dominant eigenvalue $\rho$ with the stable distribution $\Phi_{\rm r}$ as the corresponding right eigenvector and 
 the vector of reproductive values $\Psi_{\rm  r}$ as the corresponding left eigenvector. 
 \item $\Psi_{\rm  r}$ is also a left eigenvector of the scaled NGM with large domain 
\begin{linenomath*} 
 $${\mathcal L}_{\rm L}^{(\rho)} := F(\rho I -T)^{-1} = \rho^{-1}F(I-\rho^{-1}T)^{-1}$$
 \end{linenomath*} 
 corresponding to eigenvalue 1. That is, 
 it is also (modulo normalisation) the vector of generation-based reproductive values when we discount fertilities and transition probabilities by $\rho^{-1}$.
\end{enumerate}

We normalise the vector of reproductive values by requiring 
\begin{linenomath*}
\begin{equation}
\Psi_{\rm r} \Phi_{\rm r} = 1
\end{equation}
\end{linenomath*} 
and obtain 
\begin{linenomath*}
\begin{equation}
\Psi_{\rm r} = \frac{\|(\rho I - T)^{-1}V\|}{U(\rho I -T)^{-2}V} U(\rho I-T)^{-1}.
\label{Psir_scalar}
\end{equation}
\end{linenomath*}

If we now observe that 
\begin{linenomath*}
 $$(I -\rho^{-1}T)^{-2} = I + 2\rho^{-1}T + 3 \rho^{-2}T^{2} + \hdots,$$ 
 \end{linenomath*}
 we can rewrite $\Psi_{\rm r}$ as 
\begin{linenomath*}
\begin{equation}
\Psi_{\rm r} = \frac{\|(I - \rho^{-1}T)^{-1}V\|}{U(I -\rho^{-1}T)^{-2}V} U(I-\rho^{-1}T)^{-1}
\label{Psir_scalar2}
\end{equation}
\end{linenomath*} 
and interpret the denominator as $\rho^{-1}$ times the expected discounted age of the parent of a newborn individual.

\begin{example}
\label{ex:leslie}
\normalfont Consider the Leslie model with three age classes, i.e. 
\begin{linenomath*}
\begin{align*}
T &= \begin{bmatrix}
0& 0 & 0  \\
t_{21} & 0 & 0 \\
0 & t_{32} & 0
\end{bmatrix} \ \ \ \mbox{and} \ \ \  F = \begin{bmatrix}
f_{1}& f_{2} & f_{3}  \\
0 & 0 & 0 \\
0 & 0 & 0
\end{bmatrix}.
\label{Leslie}
\end{align*}
\end{linenomath*}
(See Figure \ref{fig:Leslie} for the schematic representation of the model). 

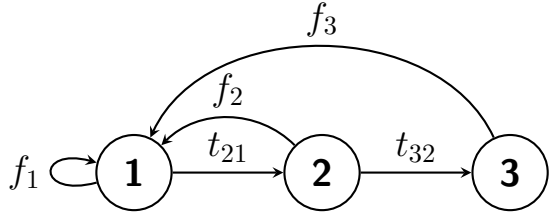
\begin{figure}[t]
\begin{center}
\begin{tikzpicture}[node distance=2.5cm and 2.5cm,font=\sffamily\Large\bfseries]
\node (N1) [state] {1};
\node (N2) [state, right of = N1] {2};
\node (N3) [state, right of = N2] {3};
\draw [arrow] (N1) -- node[anchor=south] {$t_{21}$} (N2);
\draw [arrow] (N2) -- node[anchor=south] {$t_{32}$} (N3); 
\draw[arrow] (N1) to [loop left] node [left]{$f_1$} (N1);
\draw[arrow] (N2) to [bend right = 45] node [above]{$f_2$} (N1);
\draw[arrow] (N3) to [bend right = 65] node [above]{$f_3$} (N1);
\end{tikzpicture}
\caption{A schematic representation of the Leslie model in Example \ref{ex:leslie}.}
\label{fig:Leslie}
\end{center}
\end{figure}

All individuals are born into the first age class so $F = VU$ for 
\begin{linenomath*}
\begin{equation*}
V = 
\begin{bmatrix}
1 \\ 0 \\0
\end{bmatrix} \ \ \ \mbox{and} \ \ \ U = 
\begin{bmatrix}
f_{1} & f_{2} & f_{3}
\end{bmatrix}.
\end{equation*}
\end{linenomath*}
A straightforward calculation reveals that 
\begin{linenomath*}
\begin{equation*}
(zI -T)^{-1} = 
\begin{bmatrix*}[l]
z^{-1} & 0 & 0 \\
t_{21}z^{-2} & z^{-1} & 0 \\
t_{32}t_{21}z^{-3} & t_{32}z^{-2} & z^{-1}
\end{bmatrix*},
\end{equation*}
\end{linenomath*}
which yields the Euler-Lotka equation 
\begin{linenomath*}
\begin{equation*}
 \bar{k}(z) = f_{1}z^{-1} + f_{2}t_{21}z^{-2} + f_{3}t_{32}t_{21}z^{-3} = 1.
 \end{equation*}
\end{linenomath*}
Hence 
\begin{linenomath*}
\begin{align*}
\begin{split}
1 &= f_{1}\rho^{-1} + f_{2}t_{21}\rho^{-2} + f_{3}t_{32}t_{21}\rho^{-3}, \\
{\mathcal R}_0 &= f_{1} + f_{2}t_{21} + f_{3}t_{32}t_{21}.
\end{split}
\end{align*}
\end{linenomath*}
Using \eqref{Phir_scalar_2}, we obtain the asymptotic age distribution 
\begin{linenomath*}
\begin{equation*}
\Phi_{\rm r} = \frac{1}{1 + t_{21}\rho^{-1} + t_{32}t_{21}\rho^{-2}}
\begin{bmatrix*}[c]
1 \\ t_{21}\rho^{-1} \\ t_{32}t_{21}\rho^{-2}
\end{bmatrix*},
\end{equation*}
\end{linenomath*}
while \eqref{Psir_scalar2} gives the row vector of reproductive values 
\begin{linenomath*}
\begin{align*}
\begin{split}
\Psi_{\rm r} &= c \begin{bmatrix*}[c]
f_{1}\rho^{-1} + f_{2}t_{21}\rho^{-2} + f_{3}t_{32}t_{21}\rho^{-3}, & 
 f_{3}t_{32}\rho^{-2} + f_{2}\rho^{-1}, & 
 f_{3}\rho^{-1}
\end{bmatrix*}  \\ &= 
c\begin{bmatrix*}[c]
1, &
 f_{3}t_{32}\rho^{-2} + f_{2}\rho^{-1}, &
 f_{3}\rho^{-1}
\end{bmatrix*}
\end{split}
\end{align*}
\end{linenomath*}
where the constant $c$ is determined such that $\Psi_{\rm r} \Phi_{\rm r} = 1$.

For the specific case with $t_{21} = 0.7, t_{32} = 0.8, f_1 = 0, f_2 = 1 = f_3$ we find that $\rho = 1.1$ and ${\mathcal R}_0 = 1.26$ (see Figure \ref{fig:Leslievectors}(a)). 
The asymptotic age distribution is 
\begin{linenomath*}
\begin{equation*}
\Phi_{\rm r} = 
\begin{bmatrix*}[c]
0.476 \\ 0.303 \\ 0.221
\end{bmatrix*}, 
\end{equation*}
\end{linenomath*}
while the normalised vector of reproductive values is 
\begin{linenomath*}
\begin{align*}
\Psi_{\rm r} &=  \begin{bmatrix*}[c]
0.867, & 1.363, & 0.788
\end{bmatrix*}
\end{align*}
\end{linenomath*}
(see panels (b) and (c) of Figure \ref{fig:Leslievectors}).

\begin{figure}[t]
\begin{center}
\begin{tabular}{ccc}
\includegraphics[width = 0.3\textwidth]{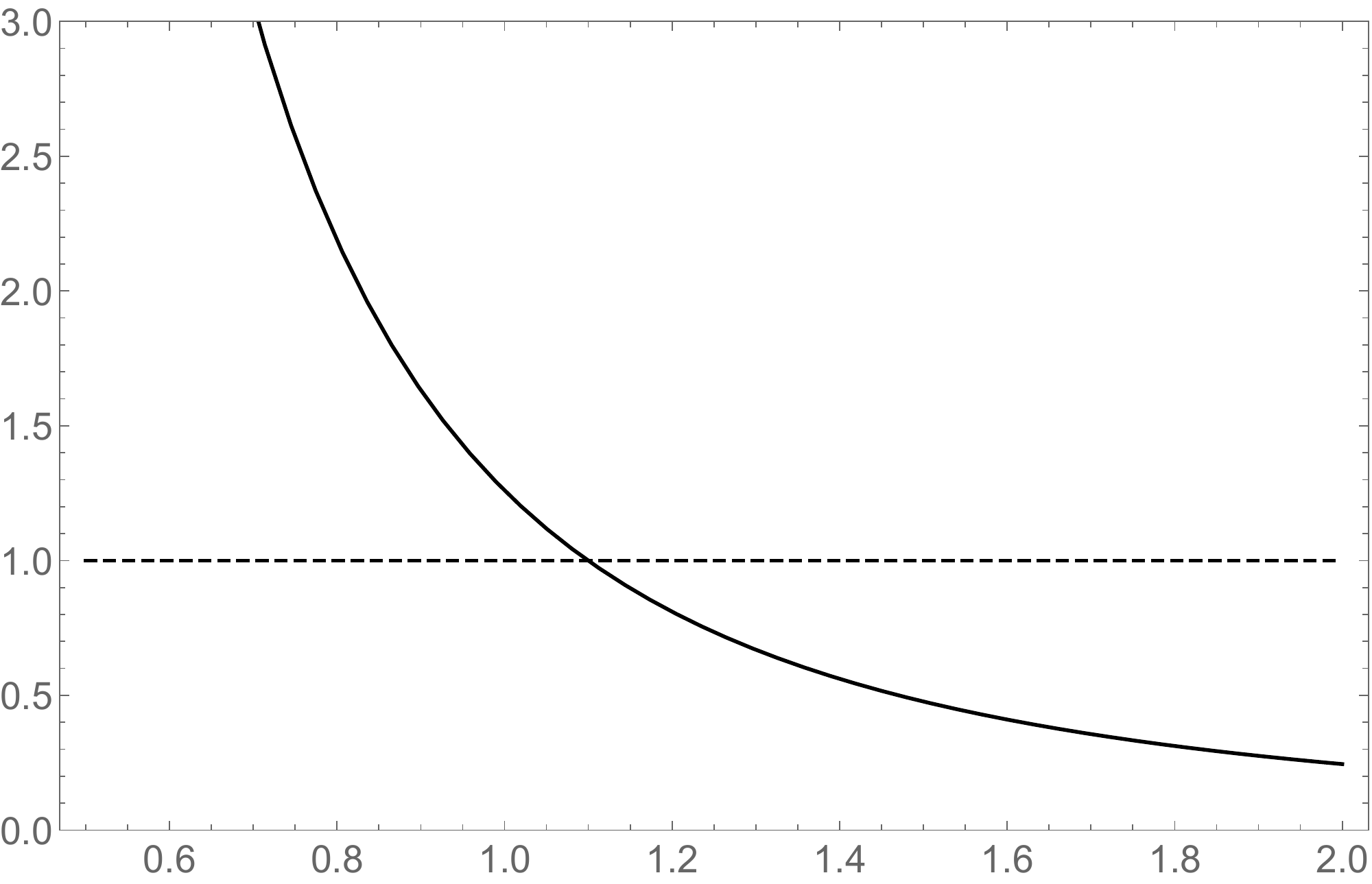} & 
\includegraphics[width = 0.3\textwidth]{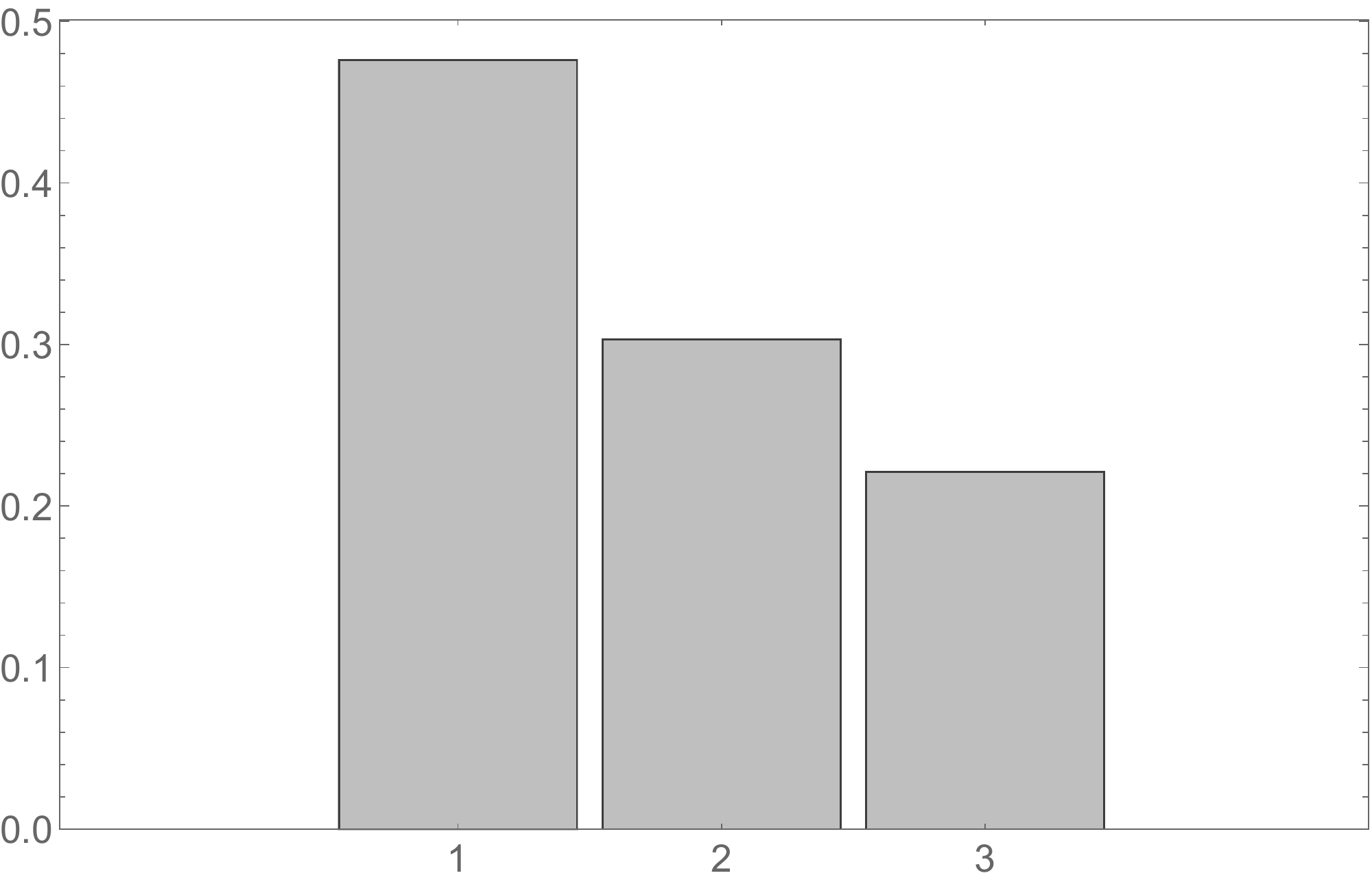} & \includegraphics[width = 0.3\textwidth]{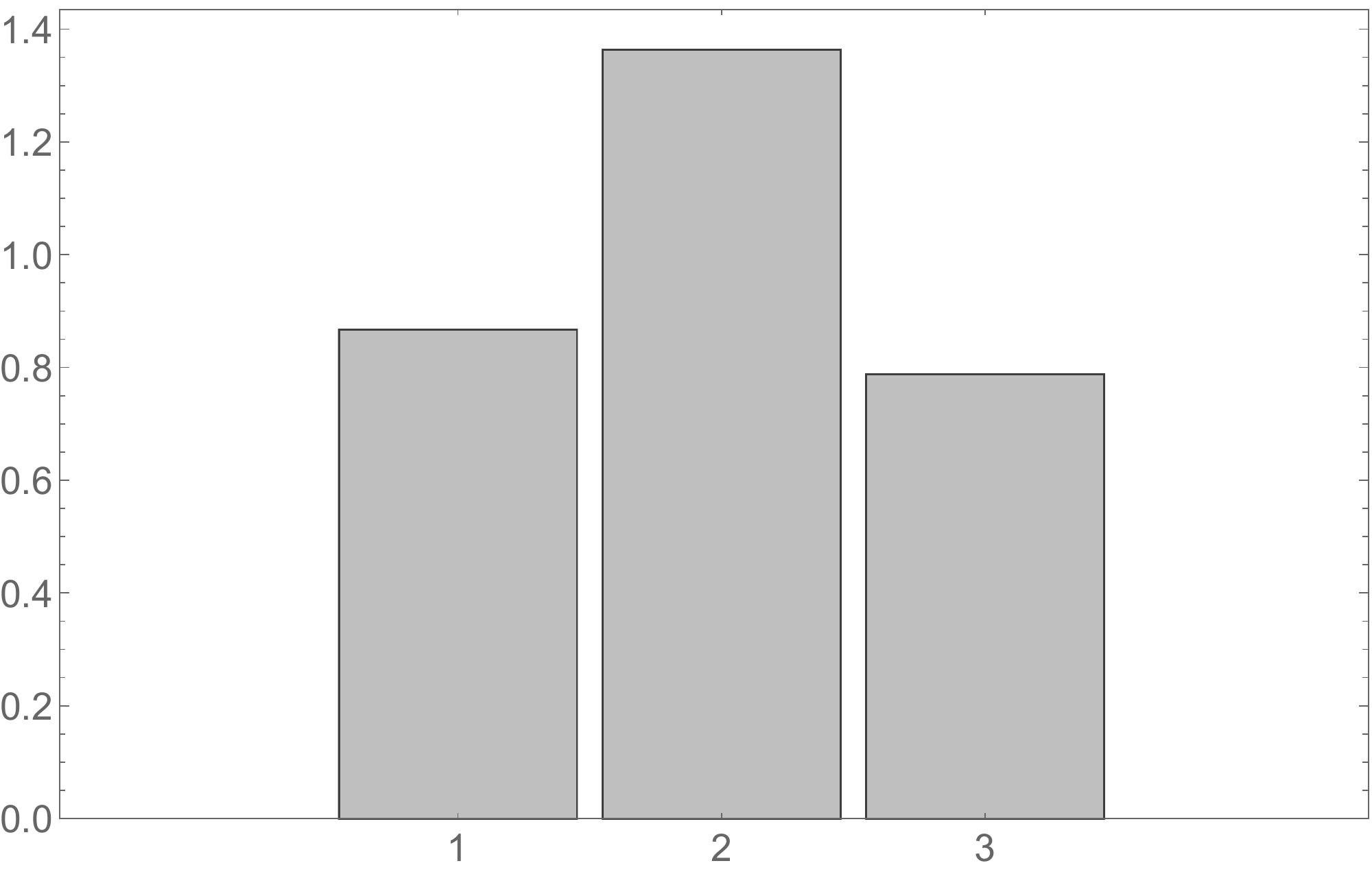}
\end{tabular}
\caption{The plots of (a) $z \mapsto \bar{k}(z)$, (b) the asymptotic age distribution and (c) the reproductive values for the (concrete example of the) Leslie model in Example \ref{ex:leslie}. }
\label{fig:Leslievectors}
\end{center}
\end{figure}

Note that if we define $N(t,a) = x_a(t)$ we can write 
\begin{linenomath*}
\begin{align*}
\begin{split}
N(t+1,a+1) &= t_{(a+1)a} N(t,a) \\
N(t+1,1) &= \sum_{a = 1}^{3}{f_a N(t,a)}
\end{split}
\end{align*}
\end{linenomath*}
which clearly exposes the Leslie matrix model as the analogue of the PDE formulation of a continuous age and time model. The RE \eqref{REbgk} is, of course, the analogue of 
Lotka's RE in the continuous age and time setting. Note also that 
$ b(t) = N(t,1).$
\end{example}

\subsection{More than one state-at-birth}

Even when individuals may be born in different $i$-states, the number of individual states-at-birth is typically significantly lower than the total number of $i$-states, thus making 
the much smaller system sizes of RE an attractive alternative to \eqref{recurrsion}. 

Suppose there are $m$ states-at-birth with $1 < m < n$ (this can now be understood literally, or, if aiming for the minimal RE system size, in a stochastic sense). 
We then write $F=VU$ for some positive $n\times m$ matrix $V$ (normalised such that all $m$ columns of 
$V$ have $l_1$-norm equal to one) and a positive $m\times n$ matrix $U$ and consider 
the RE \eqref{REBKGsys} with $\{K(s)\}_{s = 1}^{\infty}$ and $\{G(s)\}_{s = 1}^{\infty}$ as in \eqref{KGsys}. Next, require that the assumptions of Theorem \ref{TRT} hold. 

If $r_{\sigma}(T)<1$ the asymptotic dynamics of \eqref{Bsys} is completely determined by the second term. Furthermore, 
\begin{linenomath*}
\begin{equation}
\bar{K}(z) = U(zI-T)^{-1}V
\label{barKU}
\end{equation}
\end{linenomath*}
for $|z|\geq 1$. In particular, $\bar{K}(1)$ is the next-generation matrix (possibly with small domain) and 
\begin{linenomath*}
\begin{equation}
{\mathcal R}_0 = r_{\sigma}(U(I-T)^{-1}V).
\label{R0Usys}
\end{equation}
\end{linenomath*}
When ${\mathcal R}_0<1$ 
we have $B(t) \to 0$ as $t \to \infty$ and consequently $X(t)\to 0$ as well. 

Suppose  now that  ${\mathcal R}_0 \geq 1$.  The real time growth rate $\rho$  is determined as the unique solution of the Euler-Lotka equation
\begin{linenomath*}
$$r_{\sigma}(\bar{K}(z)) =  1,$$
\end{linenomath*}
where again either $\rho = 1 = {\mathcal R}_0$ or $1< \rho < {\mathcal R}_0$.
Note that rewriting \eqref{barKU} as
\begin{linenomath*}
\begin{equation}
\bar{K}(z) = z^{-1}U(I -z^{-1}T)^{-1} V,
\label{Kbarscaled}
\end{equation}
\end{linenomath*}
we can once more observe that the population growth rate $\rho$ is the value of $z$ for which both the generation and the real time process become stationary when we multiply all fertilities and 
all transition probabilities by $z^{-1}$ (and when $\rho >1$ the latter can be interpreted as introducing an additional, state-independent, death probability per time step 
$1-\rho^{-1}$). Note also that the matrix $\bar{K}(\rho)$ is the NGM (or NGM with small domain) when we discount fertilities and transition rates by a factor $\rho^{-1}$, 
\begin{linenomath*}
\begin{equation*}
\bar{K}(\rho) =  \rho^{-1}U(I -\rho^{-1}T)^{-1} V. 
\end{equation*}
\end{linenomath*}

The $m$-vector $B(t)$ will for large $t$ grow geometrically with rate $\rho$ while converging to the 
distribution $\Phi$, where $\Phi$ is the normalised positive right eigenvector of $\bar{K}(\rho)$ corresponding to eigenvalue 1. 
We can then deduce from \eqref{xy} and \eqref{YVB} that, for ${\mathcal R}_0 >1$ and a non-trivial initial condition 
$X(0) > 0$, the vector $X(t)$ grows geometrically with rate 
$\rho$ while converging to the asymptotic distribution 
\begin{linenomath*}
\begin{subequations}
\begin{align}
\Phi_{\rm r} &= \frac{1}{\|(\rho I -T)^{-1}V\Phi\|}(\rho I -T)^{-1}V\Phi \\
&= \frac{1}{\|(I -\rho^{-1}T)^{-1}V\Phi\|}(I -\rho^{-1}T)^{-1}V\Phi. 
\label{Phirsys}
\end{align}
\end{subequations}
\end{linenomath*}

When using the $l_1$ norm, we can interpret the denominator in  \eqref{Phirsys} as the expected duration of life of a newborn when newborns are sampled from 
the asymptotic distribution $\Phi$ and with additional, state-independent, death probability $1-\rho^{-1}$ introduced into the system (or alternatively, when we 
discount again and again the next year by $\rho^{-1}$).

To see how the components of $X(0)$ contribute to future population growth (i.e. to the constant $c$ in \eqref{ThiemeC}) we compute 
\begin{linenomath*}
\begin{equation}
 \bar{G}(\rho) = \rho U (\rho I-T)^{-1}.
\end{equation}
\end{linenomath*}
Using \eqref{ThiemeC} we now conclude that, if $\Psi$ is a positive left eigenvector of $\bar{K}(\rho)$ corresponding to eigenvalue 1 then 
the $j$-th element of the row $n$-vector 
\begin{linenomath*}
\begin{equation}
\Psi_{\rm r} = c\Psi U(\rho I-T)^{-1}
\label{Psirsysc}
\end{equation}
\end{linenomath*}
specifies the (relative) contribution of individuals with $i$-state $j$ to future population growth. Again, we normalise this vector by requiring 
$\Psi_{\rm r} \Phi_{\rm r} = 1$. Thus we obtain 
\begin{linenomath*}
\begin{subequations}
\begin{align}
\Psi_{\rm r} &= \frac{\|(\rho I - T)^{-1}V\Phi\|}{\Psi U(\rho I -T)^{-2}V \Phi} \Psi U(\rho I-T)^{-1} \\ 
&= \frac{\|( I - \rho^{-1}T)^{-1}V\Phi\|}{\Psi U(I -\rho^{-1}T)^{-2}V \Phi} \Psi U(I-\rho^{-1}T)^{-1}. 
\label{Psirsys}
\end{align}
\end{subequations}
\end{linenomath*}
With the same reasoning as in the previous subsection, we can now interpret the $j$-th component of $U(I -\rho^{-1}T)^{-2}V \Phi$ in the denominator of 
\eqref{Psirsys} as $\rho^{-1}$ times the expected age of the parent 
of a newborn individual with $i$-state $j$ when we sample parents according to $\Phi$ and take into account an additional, state-independent, 
death probability $1-\rho^{-1}$. 

Again, one easily verifies that 
\begin{enumerate}[(i)]
\item $F + T$ has dominant eigenvalue $\rho$ with the stable distribution $\Phi_{\rm r}$ as the corresponding right eigenvector and 
 the vector of reproductive values $\Psi_{\rm  r}$ as the corresponding left eigenvector. 
 \item $\Psi_{\rm  r}$ is also a left eigenvector of the scaled NGM with large domain
\begin{linenomath*} 
 $${\mathcal L}_{\rm L}^{(\rho)} := F(\rho I -T)^{-1} = \rho^{-1}F(I-\rho^{-1}T)^{-1}$$
 \end{linenomath*} 
 corresponding to eigenvalue 1. That is, 
 it is also the vector of (generation-based) reproductive values when we discount fertilities and transition probabilities by $\rho^{-1}$.
\end{enumerate}

Note that the conditions $A6$ and $A7$ imposed on the resolvent can be interpreted biologically. Indeed, 
by definition 
the kernel $K  = \{K(t)\}_{t = 1}^{\infty}$ contains information about the expected numbers of daughters produced by one female at various ages. The matrix $K^{2*} = K\ast K$ yields  
the expected numbers of granddaughters (i.e. second generation offspring) and the $j$-th convolution of $K$ with itself, $K^{j*}$,
yields the expected number of $j$-th generation offspring. Summing over all generations of offspring we obtain the clan kernel, 
\begin{linenomath*}
\begin{equation*}
K^{(c)} := \sum_{j =1}^{\infty}{K^{j*}}.
\end{equation*}
\end{linenomath*}
Since clan members of every female are either her daughters or clan members of one of her daughters, or, alternatively, either her daughters or daughters of a 
member of the clan, we must have 
\begin{linenomath*}
\begin{equation*} 
K^{(c)} = K + K\ast K^{(c)} = K + K^{(c)}\ast K. 
\end{equation*}
\end{linenomath*}
That is, the clan kernel $K^{(c)}$ is the resolvent of $K$.
The assumptions $A6$ and $A7$ can therefore be interpreted as follows: 
\begin{enumerate}
\item [A6.] Consider a newborn individual. At some time $t_0$ after her birth, clan members with any state-at-birth will be born, irrespective of the birth state 
of the focus individual. 
\item [A7.] Consider two newborn individuals. There exist $t_1, t_2$ with greatest common divisor one, such that it is possible to choose the birth states of the focus individuals so that at times $t_1$ and $t_2$ after their birth, clan members of at least one of them will be born.
\end{enumerate}

As a summary of the results, we provide at the end of this Section an algorithm for deducing the large time behaviour of (discrete time) structured-population models (SPM)  via the renewal 
equation formulation (and for an application see the example below). 

\begin{example}
\label{ex:agesize}
\normalfont
Consider a population in which individuals are characterised by both age $a$ and some indicator of size $s$ (for example, both age and size are important determinants of population dynamics  
of \textit{Rhododendron maximum} shoots, see 
\cite{mcgraw1989, caswell2000}) and let 
\begin{linenomath*}
$$ \{(a,s): a \in \{1,2,3\},  s \in \{1,2\}\}$$
\end{linenomath*}
be the set of $i$-states. 
We enumerate the $i$-states in the following way: 
\begin{linenomath*}
$$\begin{tabular}{|l||c|c|c|c|c|c|}
\hline
(a,s) & (1,1) & (2,1) & (3,1) & (1,2) & (2,2) & (3,2) \\ \hline
label & 1 & 2 & 3 & 4 & 5 & 6 \\
\hline
\end{tabular}.$$
\end{linenomath*}
Let  
\begin{linenomath*}
\begin{align}
T = \begin{bmatrix}
0 & 0 & 0 & 0 & 0 & 0 \\
t_{21} & 0 & 0 & 0 & 0 & 0 \\
0 & t_{32}& 0 & 0 & 0 & 0 \\
0 & 0 & 0 & 0 & 0 & 0 \\
t_{51} & 0 & 0 & t_{54} & 0 & 0 \\
0 & t_{62} & 0 & 0 & t_{65} & 0
\end{bmatrix}  \ \ \ \mbox{and} \ \ \ 
F = \begin{bmatrix}
0 & f_{12} & f_{13} & 0 & f_{15} & f_{16} \\
0  & 0 & 0 & 0 & 0 & 0 \\
0 & 0 & 0 & 0 & 0 & 0 \\
0  & 0 & 0 & 0 & f_{45} & f_{46} \\
0 & 0 & 0 & 0 & 0 & 0 \\
0 & 0 & 0 & 0 & 0 & 0
\end{bmatrix} 
\label{TFagesize}
\end{align}
\end{linenomath*}
describe the survival $\&$ state transitions and fertility rates, respectively (see Figure \ref{fig:as} for a schematic representation of the model).

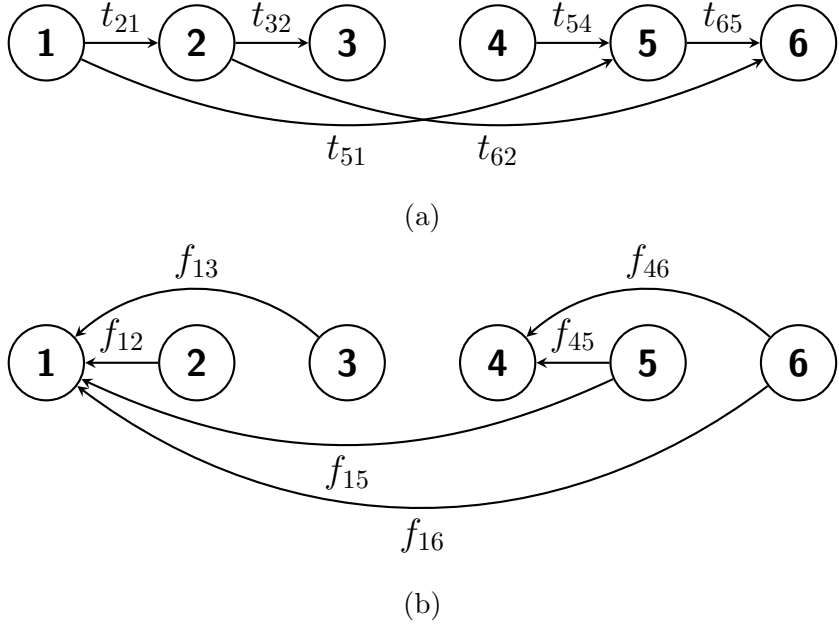
\begin{figure}[t]
\begin{center}
\begin{tikzpicture}[node distance=2.cm and 2.cm,font=\sffamily\Large\bfseries]
\node (N1) [state] {1};
\node (N2) [state, right of = N1] {2};
\node (N3) [state, right of = N2] {3};
\node (N4) [state, right of = N3] {4};
\node (N5) [state, right of = N4] {5};
\node (N6) [state, right of = N5] {6};
\draw [arrow] (N1) -- node[anchor=south] {$t_{21}$} (N2);
\draw [arrow] (N2) -- node[anchor=south] {$t_{32}$} (N3); 
\draw [arrow] (N4) -- node[anchor=south] {$t_{54}$} (N5); 
\draw [arrow] (N5) -- node[anchor=south] {$t_{65}$} (N6); 
\draw[arrow] (N1) to [bend right = 25] node [below]{$t_{51}$} (N5);
\draw[arrow] (N2) to [bend right = 25] node [below]{$t_{62}$} (N6);
\end{tikzpicture}
\\[2mm] (a) \\
\begin{tikzpicture}[node distance=2.cm and 2.cm,font=\sffamily\Large\bfseries]
\node (N1) [state] {1};
\node (N2) [state, right of = N1] {2};
\node (N3) [state, right of = N2] {3};
\node (N4) [state, right of = N3] {4};
\node (N5) [state, right of = N4] {5};
\node (N6) [state, right of = N5] {6};
\draw [arrow] (N2) -- node[anchor=south] {$f_{12}$} (N1);
\draw[arrow] (N3) to [bend right = 42] node [above]{$f_{13}$} (N1);
\draw [arrow] (N5) -- node[anchor=south] {$f_{45}$} (N4); 
\draw[arrow] (N6) to [bend right = 42] node [above]{$f_{46}$} (N4);
\draw[arrow] (N5) to [bend right = -25] node [below]{$f_{15}$} (N1);
\draw[arrow] (N6) to [bend right = -37] node [below]{$f_{16}$} (N1);
\end{tikzpicture} 
\\[2mm]
(b)
\caption{A schematic representation of (a) the state transitions  and (b) fertilities in the model described in Example \ref{ex:agesize}.}
\label{fig:as}
\end{center}
\end{figure}
There are two states at birth, i.e. $(1,1)$ and $(1,2)$, corresponding to labels 1 and 4, respectively. Then $F = VU$ for 
\begin{linenomath*}
\begin{align}
V = \begin{bmatrix}
1 & 0 \\
0 & 0 \\
0 & 0 \\
0 & 1 \\
0 & 0 \\
0 & 0 
\end{bmatrix} \ \ \  \mbox{and} \ \ \ U = 
 \begin{bmatrix}
0 & f_{12} & f_{13} & 0 & f_{15} & f_{16} \\
0  & 0 & 0 & 0 & f_{45} & f_{46} \\
\end{bmatrix}. 
\label{VUagesize}
\end{align}
\end{linenomath*}
The next-generation matrix is 
\begin{linenomath*}
\begin{align*}
{\mathcal L} &= U(I-T)^{-1}V \\&= 
 \begin{bmatrix}
f_{12}t_{21} + f_{13}t_{32}t_{21} + f_{15}t_{51} + f_{16}(t_{62}t_{21}+ t_{65}t_{51}) & f_{15}t_{54} + f_{16}t_{65}t_{54}\\
f_{45}t_{51} + f_{46}(t_{62}t_{21} + t_{65}t_{51}) & f_{45}t_{54} + f_{46}t_{65}t_{54}
\end{bmatrix}
\end{align*}
\end{linenomath*}
and ${\mathcal R}_0$ is the dominant eigenvalue of  ${\mathcal L}.$ 

Let us first consider the case where all the $t'$s and the $f'$s are strictly positive. 
Using \eqref{Ksys} we find 
\begin{linenomath*}
\begin{align*}
 K(1) &= 0, \\
K(2) &= \begin{bmatrix}
f_{12}t_{21} + f_{15}f_{51} & f_{15}t_{54} \\ f_{45}t_{51} & f_{45}t_{54}
\end{bmatrix} \gg 0 \\
K(3) &= \begin{bmatrix}
f_{13}t_{32}t_{21} + f_{16}(t_{62}t_{21} + t_{65}t_{51}) & f_{16}t_{65}t_{54}  \\ f_{46}(t_{62}t_{21} + t_{65}t_{51}) & f_{46}t_{65}t_{54}
\end{bmatrix}  \gg 0 \\
K(t) &= 0 \ \ \text{for} \ \   t \geq 4 
 \end{align*}
 \end{linenomath*}
and using \eqref{Gsys} that $G(t) = 0$ for $t \geq 3$. Therefore 
\begin{linenomath*}
\begin{align*}
R(1) &= K(1)  = 0, \\
R(2) &= K(2) \gg 0 \\
R(3) &= K(3)  \gg 0\\
R(t) &\geq K(2)R(t-2) \gg 0 \ \ \text{for} \ \   t \geq 4.
 \end{align*}
 \end{linenomath*}
The positivity of the resolvent for $t \geq 2$ can also be deduced by interpretation. Indeed, if all the $t$'s and the $f$'s are strictly positive then individuals 
with either birth state will produce offspring of either size at ages 2 and 3. For $t\geq 2$ there will therefore be born clan members with any state-at-birth, 
irrespective of the birth state of the mother. 

All the assumptions of Theorem \ref{TRT} are therefore satisfied. 
We have  
\begin{linenomath*}
$$ \bar{K}(z) = \sum_{s = 1}^{\infty}{K(s)z^{-s}} = K(2)z^{-2} + K(3)z^{-3}.$$
\end{linenomath*}
Let's assume that ${\mathcal R}_0 >1$. 
We can then compute the growth rate $\rho>1$ as the unique real solution $z$ of the Euler-Lotka equation 
\begin{linenomath*}
$$r_{\sigma}(\bar{K}(z)) = 1.$$
\end{linenomath*}
As a concrete example consider 
\begin{linenomath*}
$$T = \begin{bmatrix}
0 & 0 & 0 & 0 & 0 & 0 \\
0.3 & 0 & 0 & 0 & 0 & 0 \\
0 & 0.3& 0 & 0 & 0 & 0 \\
0 & 0 & 0 & 0 & 0 & 0 \\
0.3 & 0 & 0 & 0.5 & 0 & 0 \\
0 & 0.3& 0 & 0 & 0.5 & 0
\end{bmatrix},  \ \ \ 
F = \begin{bmatrix}
0 & 1& 1 & 0 & 1 & 1 \\
0  & 0 & 0 & 0 & 0 & 0 \\
0 & 0 & 0 & 0 & 0 & 0 \\
0  & 0 & 0 & 0 & 1 & 1 \\
0 & 0 & 0 & 0 & 0 & 0 \\
0 & 0 & 0 & 0 & 0 & 0
\end{bmatrix}. $$
\end{linenomath*}
Then 
\begin{linenomath*}
$$ \bar{K}(z) = 
\begin{bmatrix}
0.6z^{-2} + 0.33z^{-3} & 0.5z^{-2} + 0.25z^{-3} \\
0.3z^{-2} + 0.24z^{-3} & 0.5z^{-2} + 0.25z^{-3}
\end{bmatrix}.
$$
\end{linenomath*}
We can find the solution of the Euler-Lotka equation 
numerically and get $\rho = 1.18$, while ${\mathcal R}_0 = 1.48$ (see Figure \ref{fig:agesize}(a)). 

\begin{figure}
\begin{center}
\begin{tabular}{ccc}
\includegraphics[width = 0.3\textwidth]{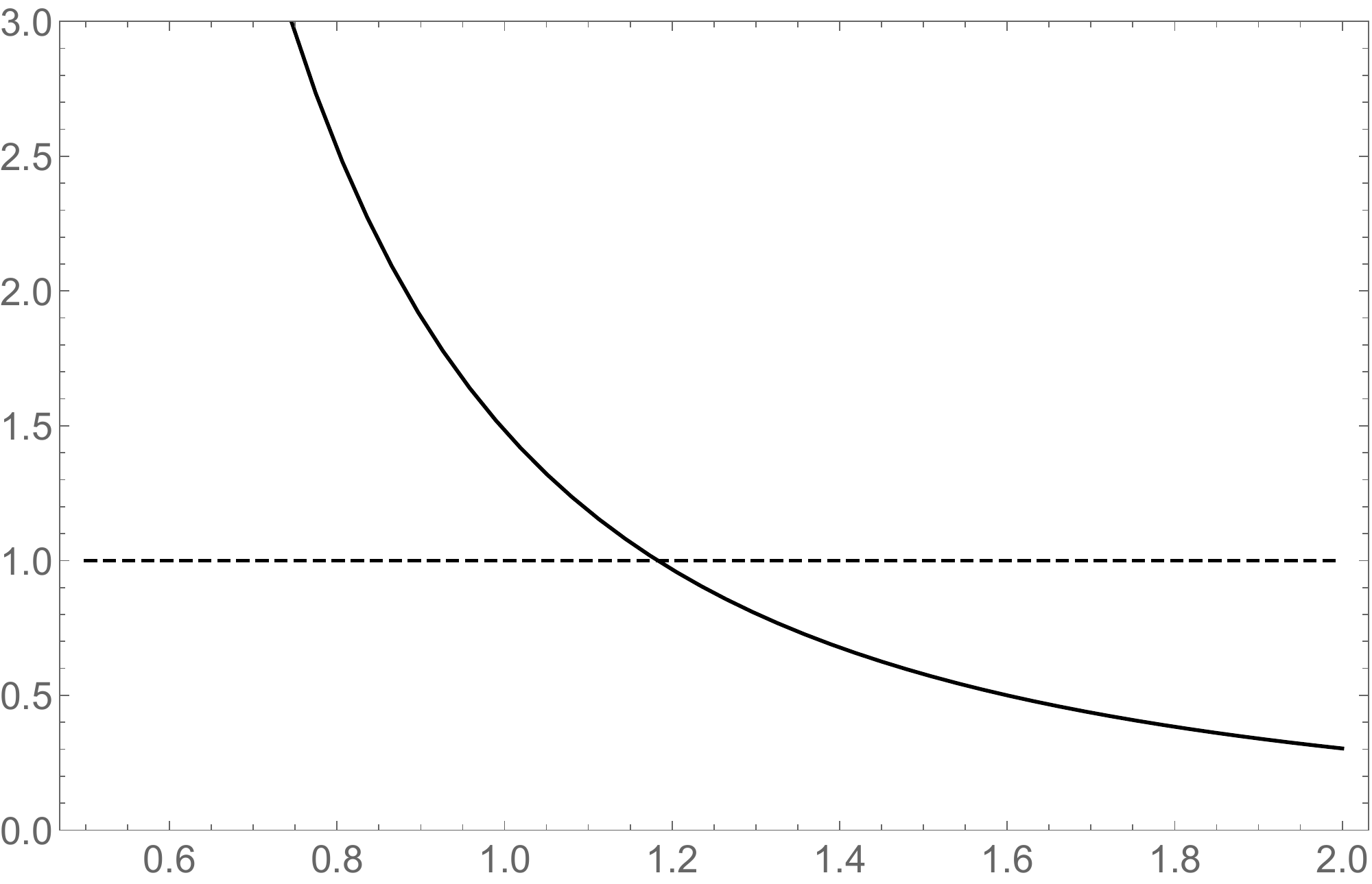} & \includegraphics[width = 0.3\textwidth]{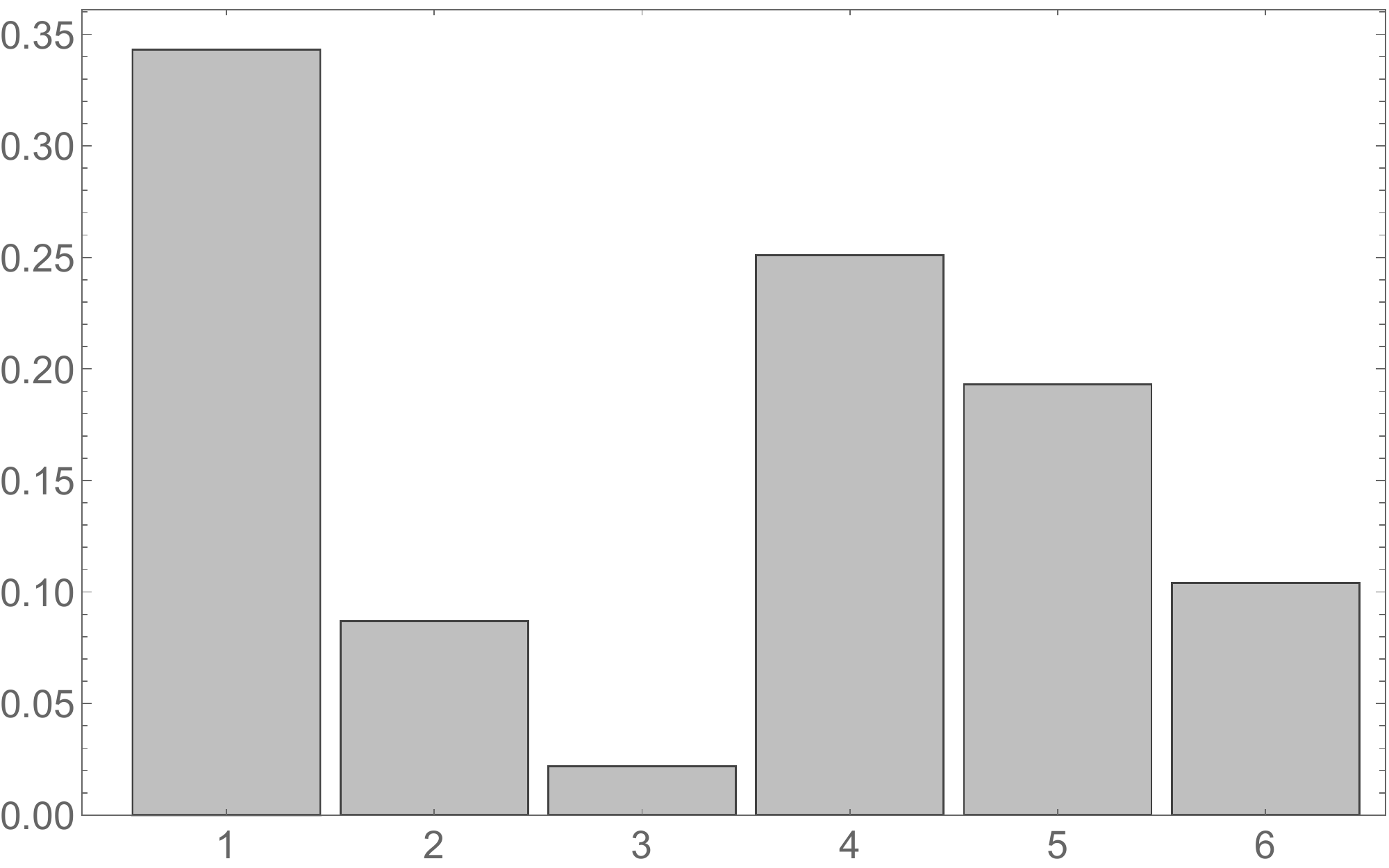} & \includegraphics[width = 0.3\textwidth]{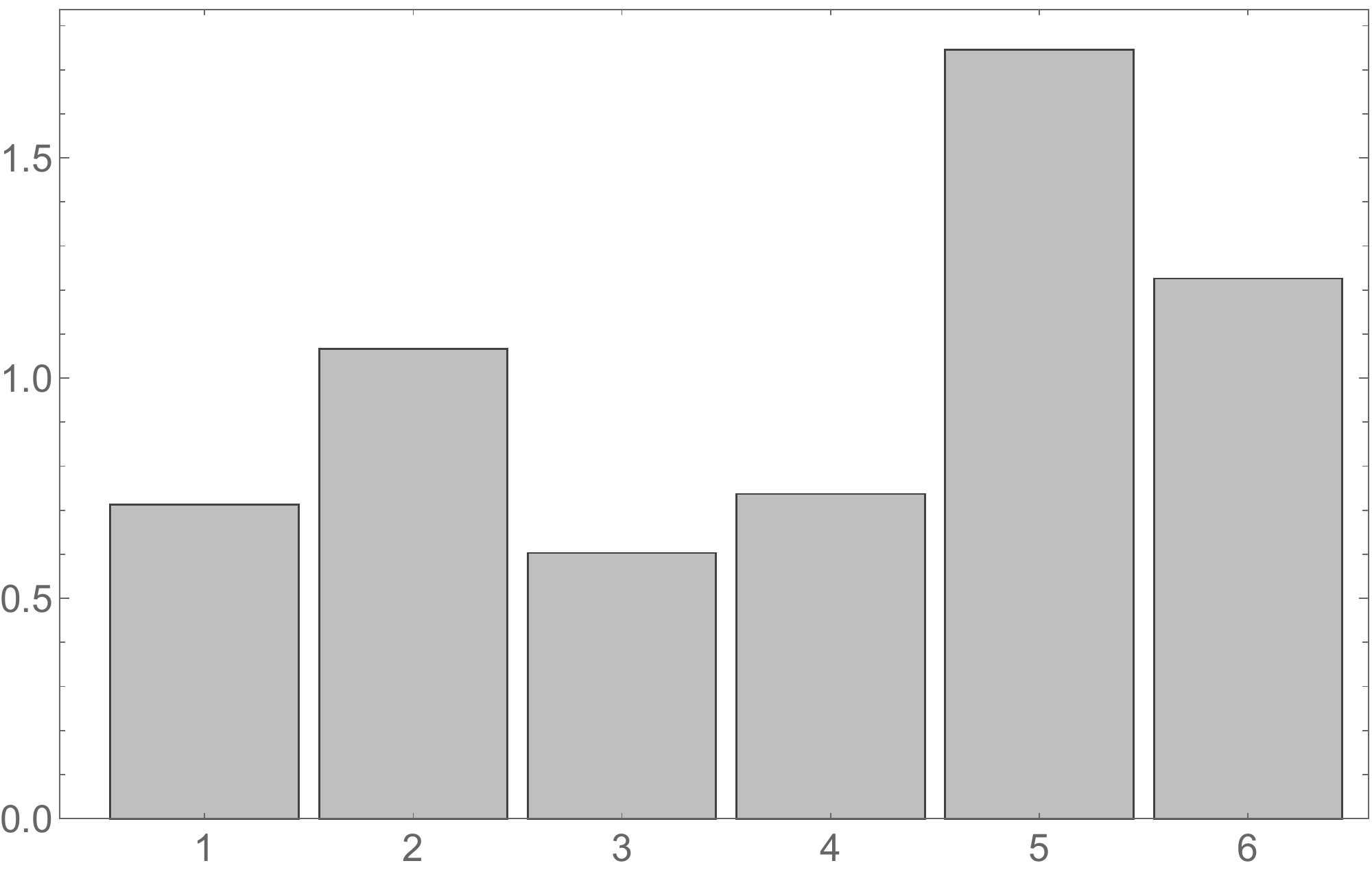} 
\end{tabular}
\end{center}
\caption{The plots of (a) $z \mapsto \textrm{spectral radius of}\ \bar{K}(z)$, (b) the asymptotic $i$-state distribution and (c) the reproductive values for (the specific example of) 
the age-size structured-population model in Example \ref{ex:agesize}.}
\label{fig:agesize}
\end{figure}

The asymptotic distribution of states-at-birth is found as the (normalised) positive eigenvector of $\bar{K}(\rho)$ corresponding to eigenvalue 1. We obtain 
\begin{linenomath*}
$$\Phi = \begin{bmatrix}
0.578 \\ 0.422
\end{bmatrix}
$$
\end{linenomath*}
and then from \eqref{Phirsys} conclude that the asymptotic population distribution is given by 
\begin{linenomath*}
$$ \Phi_{\rm r} = \begin{bmatrix}
0.343\\ 0.087\\ 0.022\\ 0.251\\ 0.193\\ 0.104
\end{bmatrix}.
$$
\end{linenomath*}
We furthermore find that the left eigenvector of $\bar{K}(\rho)$ corresponding to eigenvalue 1 (normalised such that $\Psi \Phi = 1$)  is 
\begin{linenomath*}
$$\Psi = [0.986,1.02],$$
\end{linenomath*}
from which we conclude using \eqref{Psirsys} that the normalised vector of reproductive values is 
\begin{linenomath*}
$$ \Psi_{\rm r} = \begin{bmatrix}
0.713 & 1.067 & 0.603 & 0.737 & 1.745 & 1.226 
\end{bmatrix}
$$
\end{linenomath*}
(see also panels (b) and (c) of Figure \ref{fig:agesize}).

Note that the condition of  strict positivity of $t$'s and $f$'s can be relaxed somewhat. For example, 
if we assume that 
(i) individuals of size 1 only reproduce at age 3 (and then only produce offspring of size 1) and (ii) individuals of size 2 reproduce 
at age 2 (then producing size 1 individuals) and 3 (when they produce offspring of either size) then it is clear that $R(t)$ is strictly positive for 
$t \geq 3$. 

In some situations, a further reduction of the NGM and the RE system is possible. Such is the case when we assume that (i) 
small individuals do not reproduce (i.e. $f_{12} = 0 = f_{13}$) and that (ii) large individuals produce size 1 and size 2 offspring at a given ratio $p:(1-p)$, 
regardless of the age of the parent. That is,  $f_{15} = pf_{5}, f_{16} = pf_{6}$ and 
$f_{45} = (1-p)f_{5}, f_{46} = (1-p)f_{6}$ for some $0 < p < 1$ and $f_5, f_6>0$, leading to 
\begin{linenomath*}
$$F = \begin{bmatrix}
0 & 0 & 0 & 0 & pf_{5} & pf_{6} \\
0  & 0 & 0 & 0 & 0 & 0 \\
0 & 0 & 0 & 0 & 0 & 0 \\
0  & 0 & 0 & 0 & (1-p)f_{5} & (1-p)f_{6} \\
0 & 0 & 0 & 0 & 0 & 0 \\
0 & 0 & 0 & 0 & 0 & 0
\end{bmatrix}. 
$$
\end{linenomath*}

In this case, $F$ (or $U$ in \eqref{VUagesize}) has rank 1. There are still two states-at-birth in the literal sense. However, if we 
interpret the term state-at-birth in the stochastic sense then all individuals are identical at birth in the sense that they are all born with $i$-state distribution 
\begin{linenomath*}
$$
V = \begin{bmatrix}
p \\
0  \\
0  \\
1-p \\
 0 \\
 0 
\end{bmatrix}.
$$
\end{linenomath*}
Then $F = VU$ for 
\begin{linenomath*}
$$
U = 
 \begin{bmatrix}
0 & 0 & 0 & 0 & f_{5} & f_{6} 
\end{bmatrix}$$
\end{linenomath*}
and  the NGM with the small domain amounts to the scalar 
\begin{linenomath*}
$$ {\mathcal L}_{\rm S} = U(I-T)^{-1}V = p(f_{5}t_{51} + f_{6}(t_{62}t_{21} + t_{65}t_{51})) + (1-p)(f_5 t_{54} + f_{6}t_{65}t_{54}) = {\mathcal R}_0.$$
\end{linenomath*}
In this special case, the basic reproduction number ${\mathcal R}_0$ can also easily be deduced from interpretation (and we invite the reader to do so). 

We now have 
\begin{linenomath*}
\begin{align*}
 k(1) &= 0, \\
k(2) &= p f_5 t_{51} + (1-p)f_5 t_{54} > 0 \\
k(3) &= pf_6(t_{62}t_{21}+t_{65}t_{51}) + (1-p)f_6t_{65}t_{54} \\
k(t) &= 0 \ \ \text{for} \ \   t \geq 4 
 \end{align*}
 \end{linenomath*}
and the scaler Euler-Lotka equation takes the form 
\begin{linenomath*}
\begin{equation}
\bar{k}(z) = \sum_{s = 1}^{\infty}{k(s)} = k(2)z^{-2} + k(3)z^{-3} = 1.
\label{kagesize}
\end{equation}
\end{linenomath*}
If ${\mathcal R}_0>1$ we can then first find the growth rate $\rho$ as the unique real solution of the Euler-Lotka equation \eqref{kagesize} and 
then use \eqref{Phir_scalar_2} and \eqref{Psir_scalar2} to determine the asymptotic $i$-state distribution and the asymptotic 
reproductive values, respectively.

\end{example}
\setlength{\fboxsep}{10pt}
\setlength{\fboxrule}{0.5pt}
\fbox{
\begin{minipage}{0.95\textwidth}{
{\sc Deducing the asymptotic behaviour in (linear discrete time) SPM from the corresponding RE}

\vspace{2mm}
{\small 
Starting with a structured-population model 
$$ X(t+1) = (F+T)X(t), \ \ \ X(t)\in {\R}^n,$$
the recipe is as follows.
\begin{enumerate}[\hspace{5pt}{\sc Step} 1.]
\item Identify states-at-birth and write the fertility matrix in the form $F = VU$ for some normalised $n\times m$ birth-state matrix $V$ and the $m\times n$ 
fertility matrix $U$. If $V$ has rank 1 proceed with the (a) part of the algorithm, otherwise follow the (b) part. 

{\em Remark.} For practical purposes, it may be handy to order the $i$-states such that the birth states come first. Then in the case of $m$ states-at-birth 
(in the literal sense)  $V$ is the matrix with columns $e_{1}, \hdots, e_{m}$ and  $U$ is obtained by taking 
the rows $1, \hdots, m$ from $F$. 

\item 
\begin{enumerate}[(a)]
\item Derive $k(t)$ and $g(t)$ using \eqref{kg} and check that the assumptions of Theorem \ref{FRT} hold. Furthermore, check that the spectral radius of $T$ is 
below one \footnotemark.
\item Derive $K(t)$ and $G(t)$ using \eqref{KGsys} and check that the assumptions of Theorem \ref{TRT} hold. Furthermore, check that 
the spectral radius of $T$ is below one \textsuperscript{1}.
\end{enumerate}
\item \begin{enumerate}[(a)]
\item Compute ${\mathcal R}_0$ in \eqref{R0U}. If ${\mathcal R}_0<1$ conclude that the population dies out. If ${\mathcal R}_0 \geq 1$ compute the growth rate $\rho \geq 1$ as the unique real 
solution of the E-L equation \eqref{ELscalar} with $\bar{k}$ in \eqref{barkU} (with $\rho =1$ whenever ${\mathcal R}_0 = 1$). Conclude that:
\begin{itemize}
\item the population grows geometrically with rate $\rho$ and  the asymptotic $i$-state distribution is given by 
\begin{align*}
\Phi_{\rm r} = \frac{1}{\|(\rho I - T)^{-1}V\|}(\rho I- T)^{-1}V, 
\end{align*}
\item the (real time) reproductive values are collected in 
\begin{equation*}
\Psi_{\rm r} = \frac{\|(\rho I - T)^{-1}V\|}{U(\rho I -T)^{-2}V} U(\rho I-T)^{-1}.
\end{equation*}
\end{itemize}
\item Compute ${\mathcal R}_0$ in \eqref{R0Usys}. If ${\mathcal R}_0<1$ conclude that the population dies out. If ${\mathcal R}_0 \geq 1$ compute the growth rate $\rho \geq 1$ as the unique real 
solution of the E-L equation \eqref{ELsystems} with $\bar{K}$ in \eqref{barKU} (with $\rho =1$ whenever ${\mathcal R}_0 = 1$). Determine the asymptotic state-at-birth distribution $\Phi$ and the asymptotic (generation-based) reproductive values $\Psi$ 
as, respectively, the right and the left positive normalised eigenvectors of $\bar{K}(\rho)$ corresponding to eigenvalue 1. 
Conclude that:
\begin{itemize}
\item the population grows geometrically with rate $\rho$ and the  asymptotic $i$-state distribution is given by 
\begin{align*}
\Phi_{\rm r} = \frac{1}{\|(\rho I -T)^{-1}V\Phi\|}(\rho I -T)^{-1}V\Phi,  
\end{align*}
\item the  (real time) reproductive values are collected in 
\begin{equation*}
\Psi_{\rm r} = \frac{\|(\rho I - T)^{-1}V\Phi\|}{\Psi U(\rho I -T)^{-2}V \Phi} \Psi U(\rho I-T)^{-1}. 
\end{equation*}
\end{itemize}
\end{enumerate}
\end{enumerate}}
\footnotetext{\textsuperscript{1} A sufficient condition is that all column sums of $T$ are strictly below 1.}
}
\end{minipage}
}

\section{The case ${\mathcal R}_0 = 1 = \rho$}

We now show directly that in the special case when ${\mathcal R}_0 = 1 = \rho$ we can compute
\begin{enumerate}[(i)]
\item the asymptotic distribution (i.e. the right eigenvector) and 
\item the reproductive values (i.e. the left eigenvector) 
\end{enumerate}
in the real time setting from those obtained in the generation-bookkeeping framework. Note that the former result has been demonstrated in a more general setting 
in \cite{diekmann1998, diekmann2003}. 

Suppose that there are $m$ states-at-birth in the literal sense and as before write $F = VU$ for some normalised $n \times m$ birth-state matrix $V$ and the $m\times n$ 
fertility matrix $U$. Then 
${\mathcal L} = U(I-T)^{-1}V$ is the next-generation matrix. 

Let $\Phi_g$ and $\Phi_{\rm r}$ denote the normalised positive right eigenvector in, respectively, generation and real time framework, corresponding to eigenvalue 1. 
That is 
\begin{linenomath*}
\begin{subequations}
\begin{align}
U(I-T)^{-1}V\Phi_g &= \Phi_g, \ \ \  \|\Phi_g\| = 1 
\label{Phig}
\intertext{and}
(VU + T)\Phi_{\rm r} &= \Phi_{\rm r}, \ \ \ \|\Phi_{\rm r}\| = 1.
\end{align}
\end{subequations}
\end{linenomath*}
Now apply $V$ from the left on both sides of \eqref{Phig} and write 
\begin{linenomath*}
\begin{align*}
VU(I-T)^{-1}V\Phi_g &= (I-T)(I-T)^{-1}V\Phi_g.
\end{align*}
\end{linenomath*}
If we now denote $\Xi = (I-T)^{-1}V\Phi_g$ then clearly $(VU + T)\Xi = \Xi$. We thus find that the asymptotic distributions in real time and in the generation setting 
are related by 
\begin{linenomath*}
\begin{equation}
\Phi_{\rm r} = \frac{1}{\|(I-T)^{-1}V\Phi_g\|}(I-T)^{-1}V\Phi_g, 
\label{Phir}
\end{equation}
\end{linenomath*}
as indeed follows from \eqref{Phirsys} when we observe that with ${\mathcal R}_0 = 1 = \rho$, $\bar{K}(\rho)$ is the NGM 
and $\Phi = \Phi_g$. 
When using the $l_1$ norm, the denominator in \eqref{Phir} is the expected duration of life of a newborn when we sample newborns according to  $\Phi_g$.  
As observed before, 
$V\Phi_g$ is  the asymptotic distribution in the generation-bookkeeping framework when we consider the 
NGM with large domain ${\mathcal L}_{\rm L} = F(I-T)^{-1}$.

Now let $\Psi_g$ and $\Psi_{\rm r}$ denote the normalised positive 
left eigenvector in, respectively, generation and real time framework, corresponding to eigenvalue 1, i.e. 
\begin{linenomath*}
\begin{subequations}
\begin{align}
\Psi_gU(I-T)^{-1}V &=  \Psi_g, \ \ \ \Psi_g\Phi_g = 1 
\label{Psig}
\intertext{and}
\Psi_{\rm r}(VU + T) &= \Psi_{\rm r}, \ \ \ \Psi_{\rm r}\Phi_{\rm r} = 1.
\end{align}
\end{subequations}
\end{linenomath*}
Acting with $U$ from the right on both sides of \eqref{Psig} we find that 
\begin{linenomath*}
\begin{align*}
\Psi_gU(I-T)^{-1}VU &= \Psi_gU(I-T)^{-1}(I-T).
\end{align*}
\end{linenomath*}
Then for $\Xi = \Psi_gU(I-T)^{-1}$ we have $\Xi (VU + T) = \Xi$. We thus find that the reproductive values in the real time setting relate to the ones in the generation-bookkeeping 
by 
\begin{linenomath*}
\begin{equation*}
\Psi_{\rm r} = c\Psi_gU(I-T)^{-1}, 
\end{equation*}
\end{linenomath*}
where the constant $c$ is chosen such that with $\Phi_{\rm r}$ in $\eqref{Phir}$ we have $\Psi_{\rm r}\Phi_{\rm r} = 1$. We obtain 
\begin{linenomath*}
\begin{equation}
\Psi_{\rm r} = \frac{\|(I - T)^{-1}V\Phi_g\|}{\Psi_g U(I -T)^{-2}V \Phi_g} \Psi_g U(I-T)^{-1}. 
\label{Psir}
\end{equation}
\end{linenomath*}

Alternatively, we can deduce this from \eqref{Psirsys} by 
taking into account that with ${\mathcal R}_0 = 1 = \rho$ we have $\Psi = \Psi_g$. The $j$-th component of $U(I -T)^{-2}V \Phi_g$ in the denominator of 
\eqref{Psir} is 
the expected age of the parent 
of a newborn individual with $i$-state $j$ when we sample parents according to $\Phi_g$. As observed before, 
$\Psi_r$ is also the vector of generation-based reproductive values when we consider the 
NGM with large domain ${\mathcal L}_{\rm L} = F(I-T)^{-1}$.

\section{Concluding remarks}

 The specification of a linear physiologically structured population model involves two rules, one for reproduction and one for development/maturation/movement and 
 survival. Once these ingredients are specified, one can constructively define next-population-state operators. For constant environments, i.e., autonomous dynamics, a 
 first question is: will the population ultimately grow exponentially (in which case one has to ponder the issue of density dependence) or decline and go extinct? As 
 emphasised by Jim Cushing and Zhou Yicang \cite{cushing1994}, an efficient way of answering the question is to adopt a generation perspective, by focusing 
 on expected lifetime offspring production by newborn individuals, and to characterise the appropriate average as the dominant eigenvalue of the next-generation matrix.  
    On the other hand, people familiar with the Perron-Frobenius theory of positive dynamical systems (as presented in, for instance,\cite{berman1994, batkai2017}) 
    will be inclined to compute the spectral bound of the generator of the real time dynamics.
   Here, in the spirit of the elegant Li-Schneider paper \cite{li2002}, we have uncovered how these two approaches relate to each other. Our main contribution has 
   been to highlight the connecting role of the Renewal Equation and to revive the powerful results of Feller and Thieme that describe the asymptotic large time behaviour of 
   its solutions. As a one sentence summary of the present paper we offer: use the ingredients to formulate the RE, apply Feller/Thieme, and everything you might be 
   interested in can next be computed explicitly.
   
   \section*{Acknowledgements}
   Barbara Boldin acknowledges the support of  the Slovenian Research Agency (I0-0035, research program P1-0285 and research projects J1-9186, J1-1715). 
   Hans Metz benefited from the support from the “Chaire Modélisation Mathématique et Biodiversité of Veolia Environnement-École Polytechnique-Museum National d’Histoire Naturelle-Fondation X.”
   
   \appendix
   \section*{Appendix. A concise summary of the continuous time formalism}

\renewcommand{\theequation}{A.\arabic{equation}}
% redefine the command that creates the equation number.
\renewcommand{\thetable}{A.\arabic{table}}
\setcounter{equation}{0}  % reset counter 
\setcounter{figure}{0}
\setcounter{table}{0}

As the analogue of \eqref{recurrsion} we consider the linear ODE system 
\begin{linenomath*}
\begin{equation}
\frac{dX}{dt} = (F+T)X,
\label{app:sys}
\end{equation}
\end{linenomath*}
where $X$ takes values in $\R^n$, $F$ is a positive $n \times n$ matrix describing reproduction and $T$ is a positive-off-diagonal $n\times n$ matrix 
describing survival and development. We assume that the spectral bound $s(T)$ is negative, allowing us to write 
\begin{linenomath*}
\begin{equation}
-T^{-1} = \int_{0}^{\infty}{e^{aT}da}.
\end{equation}
\end{linenomath*}
The variation-of-constants version of \eqref{app:sys} reads 
\begin{linenomath*}
\begin{equation}
X(t) = e^{tT}X(0) + \int_{0}^{t}{e^{\tau T}FX(t-\tau)d\tau}.
\label{app:voc}
\end{equation}
\end{linenomath*}
Putting 
\begin{linenomath*}
\begin{equation}
Y(t) = FX(t)
\end{equation}
\end{linenomath*}
and applying $F$ to both sides of \eqref{app:voc} we obtain the RE
\begin{linenomath*}
\begin{equation}
Y(t) = Fe^{tT}X(0) + \int_{0}^{t}{Fe^{\tau T}Y(t -\tau)d\tau}
\label{app:REY}
\end{equation}
\end{linenomath*}
and, once we have constructed $Y$ by solving this RE, we can interpret \eqref{app:voc} as an explicit expression for $X$:
\begin{linenomath*}
\begin{equation}
X(t) = e^{tT}X(0) + \int_{0}^{t}{e^{\tau T}Y(t-\tau)d\tau}.
\end{equation}
\end{linenomath*}
Whenever $F$ has the form \eqref{FVU}, we can introduce $B(t) = UX(t)$. Then 
$Y(t) = VB(t)$ and we can rewrite \eqref{app:REY} 
as 
\begin{linenomath*}
\begin{equation}
B(t) = Ue^{tT}X(0) + \int_{0}^{t}{Ue^{\tau T}VB(t -\tau)d\tau}.
\label{app:REB}
\end{equation}
\end{linenomath*}

The translation invariant version of \eqref{app:REB} is 
\begin{linenomath*}
\begin{equation}
B(t) = \int_{0}^{\infty}{Ue^{\tau T}VB(t-\tau)d\tau}.
\label{app:B}
\end{equation}
\end{linenomath*}
Solutions of \eqref{app:sys} and \eqref{app:B} defined for all times, i.e. for $t \in \R$, are related to each other by 
\begin{linenomath*}
\begin{align}
\begin{split}
B(t) &= UX(t), \\
X(t) &= \int_{0}^{\infty}{e^{\tau T}VB(t -\tau)d\tau}.
\end{split}
\label{app:BX}
\end{align}
\end{linenomath*}
Clearly \eqref{app:sys} has a solution of the form 
\begin{linenomath*}
\begin{equation}
X(t) = e^{rt}\Phi_{\rm r}
\end{equation}
\end{linenomath*}
if and only if $r$ is an eigenvalue of $F+T$ and $\Phi_{\rm r}$ is a corresponding eigenvector, i.e. 
\begin{linenomath*}
\begin{equation}
(F+T)\Phi_{\rm r} = r \Phi_{\rm r}.
\end{equation}
\end{linenomath*}
By substitution we see that \eqref{app:B} has a solution of the form 
\begin{linenomath*}
\begin{equation}
B(t) = e^{rt}\Phi
\end{equation}
\end{linenomath*}
if and only if $U(rI-T)^{-1}V$ has eigenvalue 1 and $\Phi$ is a corresponding eigenvector, i.e. 
\begin{linenomath*}
\begin{equation}
\Phi = U(rI-T)^{-1}V\Phi.
\end{equation}
\end{linenomath*}
According to \eqref{app:BX} we have 
\begin{linenomath*}
\begin{align}
\begin{split}
\Phi &= U\Phi_r, \\
\Phi_r &= (rI-T)^{-1}V\Phi.
\end{split}
\label{app:PhiPhir}
\end{align}
\end{linenomath*}
We recall how this can be used in practice: 
\begin{enumerate}[\hspace{5pt}{\sc Step} 1.]
\item Find the real number $r$ for which the spectral radius of $U(rI-T)^{-1}V$ equals one, by exploiting that the spectral radius 
is a monotone decreasing function of $r$ (see \cite{heijmans1986, inaba2017, franco}).
\item Find $\Phi$ as the dominant positive eigenvector of $U(rI-T)^{-1}V$.
\item Compute $\Phi_r$ by using \eqref{app:PhiPhir}; if desired, renormalise to obtain the multiple of $\Phi_r$ that has norm one.
\end{enumerate}
For the corresponding left eigenvectors we have 
\begin{linenomath*}
\begin{align}
\begin{split}
\Psi_r(F+T) &= r\Psi_r, \\
\Psi &= \Psi F(rI-T)^{-1}.
\end{split}
\label{app:PsiPsir}
\end{align}
\end{linenomath*}
Rewriting the first of these as follows: 
\begin{linenomath*}
$$ \Psi_r F = \Psi_r (rI-T)^{-1} \iff \Psi_r F(rI-T)^{-1} = \Psi_r$$
\end{linenomath*}
we see that, modulo normalisation, $\Psi_r$ and $\Psi$ are identical. Note that the NGM with large domain is now given by 
\begin{linenomath*}
\begin{equation}
{\mathcal L}_{\rm L}:= -FT^{-1}
\end{equation}
\end{linenomath*}
and that, with ${\mathcal R}_0$ defined as the spectral radius of ${\mathcal L}$, the relation 
\begin{linenomath*}
\begin{equation}
{\rm sign} ({\mathcal R}_0-1) = {\rm sign} \  r
\end{equation}
\end{linenomath*}
holds. In the special case ${\mathcal R}_0 = 1, r=0$, $\Psi$ and $\Phi$ are, respectively, the left and the right eigenvector of ${\mathcal L}$ corresponding to its dominant 
eigenvalue. 

The formalism easily extends to an $i$-state space that is a continuum and to measures see \cite{diekmann1998, heijmans1986, thieme2009, franco}. A convenient 
starting point is the following generalisation of \eqref{app:BX}:
\begin{linenomath*}
\begin{align}
\begin{split}
b(t, \omega) &= \int_{\Omega}{\beta(\eta, \omega)x(t, d\eta)}, \\
x(t, \omega) &= \int_{0}^{\infty}{\int_{\Omega}{u(a,\xi,\omega)b(t-a,d\xi)da}}.
\end{split}
\end{align}
\end{linenomath*}
By substituting the second of these identities into the first we obtain the RE 
\begin{linenomath*}
\begin{equation}
b(t, \omega) = \int_{0}^{\infty}{\int_{\Omega}{K(a,\xi,\omega)b(t-a, d\xi)}da}
\label{app:b}
\end{equation}
\end{linenomath*}
with 
\begin{linenomath*}
\begin{equation}
K(a, \xi, \omega) := \int_{\Omega}{\beta(\eta, \omega)u(a, \xi, d\eta)}.
\end{equation}
\end{linenomath*}
Clearly, \eqref{app:b} has a solution of the form 
\begin{linenomath*}
\begin{equation}
b(t, \omega) = e^{rt} \Phi(\omega)
\end{equation}
\end{linenomath*}
if and only if 
\begin{linenomath*}
\begin{equation}
\Phi(\omega) = \int_{\Omega}{\left(\int_{0}^{\infty}{K(a, \xi, \omega)e^{-ra}da}\right)\Phi(d\xi)}.
\end{equation}
\end{linenomath*}
We refer to \cite{franco} for a detailed analysis.

\bibliographystyle{unsrt}
\bibliography{BoldinDiekmannMetz_PopGrowth}
\end{document}